\documentclass[10pt]{iopart}

\usepackage{iopams} 

\expandafter\let\csname equation*\endcsname\relax
\expandafter\let\csname endequation*\endcsname\relax
\usepackage{amsmath}

\usepackage[T1]{fontenc}
\usepackage[utf8]{inputenc} 
\usepackage{graphicx}
\usepackage[]{hyperref}
\usepackage[version=3]{mhchem}
\usepackage{subcaption}
\usepackage{amssymb}
\usepackage{gensymb}
\usepackage{longtable}
\usepackage{multirow}
\usepackage{mathtools}
\usepackage{booktabs}

\usepackage{upgreek}

\usepackage{epstopdf}
\usepackage{caption}

\usepackage{color}
\usepackage{xcolor}

\usepackage{graphicx}
\usepackage[labelformat=simple]{subcaption} 

\usepackage{placeins}
\usepackage[font={small}]{caption}

\usepackage{mathrsfs}

\usepackage{enumitem}
\usepackage{url}
\usepackage{times}

\usepackage{pgfplots}
\usepgfplotslibrary{groupplots}
\usepackage{tikz}
\usetikzlibrary{patterns}
\usetikzlibrary{external}
\usetikzlibrary{spy}

\pgfplotsset{
    colormap={violet}{rgb255=(25,25,122) color=(white) rgb255=(238,140,238)}
}

\pgfplotsset{
    colormap={vir_map}{rgb255=(254,188,43) rgb255=(219,92,104) rgb255=(13,8,135)}
}

\usepackage{pgfkeys}
\usetikzlibrary{intersections}
\usepackage{listings}

\newlength\Colsep
\newcommand{\ex}{\hat{\vec e}_{x}}
\newcommand{\ey}{\hat{\vec e}_{y}}
\newcommand{\ez}{\hat{\vec e}_{z}}

\renewcommand{\vec}[1]{\boldsymbol{#1}}
\renewcommand{\mat}[1]{\mathbf{#1}}
\newcommand{\der}[2]{\frac{\partial #1}{\partial #2}}

\newcommand{\derd}[2]{\frac{d #1}{d #2}}
\newcommand{\paren}[1]{\left(#1\right)}

\newcommand{\volInt}[3]{\paren{#1 \, , #2}_{#3}}

\newcommand{\grad}{\textrm{\textbf{grad}}\, }

\renewcommand{\div}{\textrm{div}\, }

\newcommand{\curl}{\textrm{\textbf{curl}}\, }

\renewcommand{\b}{\vec b}

\newcommand{\h}{\vec h}
\newcommand{\m}{\vec m}

\renewcommand{\e}{\vec e}

\renewcommand{\j}{\vec j}

\newcommand{\jc}{j_{\text{c}}}

\newcommand{\hs}{\vec h_{\text{app}}}
\newcommand{\hsS}{h_{\text{app}}} 
\newcommand{\hmax}{h_{\text{max}}} 
\newcommand{\hin}{\h_{\text{in}}}
\newcommand{\hinS}{h_{\text{in}}}

\renewcommand{\O}{\Omega}

\newcommand{\Om}{\Omega_{\text{m}}}

\newcommand{\Omc}{\Omega_{\text{m}}^{\text{C}}}

\newcommand{\hpf}{$h$-$\phi$-formulation\ }
\newcommand{\pf}{$\phi$-formulation\ }

\newcommand{\hpfOnly}{$h$-$\phi$-formulation}
\newcommand{\pfOnly}{$\phi$-formulation}

\newcommand{\bhyst}{\boldsymbol{\mathcal{B}}}
\newcommand{\SC}{S cell}
\newcommand{\CSC}{CS cell}
\newcommand{\SCs}{S cells}
\newcommand{\CSCs}{CS cells}

\usepackage{eso-pic}

\newcommand{\hrev}{\h_{\text{rev}}}%
\newcommand{\hrevp}{\h_{\text{rev(p)}}}%
\newcommand{\dhrev}{\dot\h_{\text{rev}}}%
\newcommand{\hrevk}{\h_{\text{rev},k}}%

\newcommand{\g}{\boldsymbol{g}}%

\newcommand{\hirr}{\h_{\text{irr}}}%
\newcommand{\hirrk}{\h_{\text{irr},k}}%
\newcommand{\hirru}{\h_{\text{irr}}}%
\newcommand{\heddy}{\h_{\text{eddy}}}%
\newcommand{\hcoupling}{\h_{\text{coupling}}}%
\newcommand{\dbcoupling}{\dot \b_{\text{coupling}}}%

\newcommand{\teddy}{\tau_{\text{e}}}%
\newcommand{\tcoupling}{\tau_{\text{c}}}%

\newcommand{\tek}{\tau_{\text{e},k}}%
\newcommand{\tck}{\tau_{\text{c},k}}%

\newcommand{\ku}{\kappa}%
\newcommand{\kc}{\chi}%

\newcommand{\ptot}{p_{\text{tot}}}%
\newcommand{\prev}{p_{\text{rev}}}%
\newcommand{\pirr}{p_{\text{irr}}}%
\newcommand{\pirru}{p_{\text{irr}}}%
\newcommand{\pirrc}{p_{\text{irr,c}}}%
\newcommand{\peddy}{p_{\text{eddy}}}%
\newcommand{\pcoupling}{p_{\text{coupling}}}%

\usepackage{color}
\usepackage{xcolor}
\definecolor{myred}{rgb}{0.7,0.15,0.15}
\definecolor{mymaincolor}{rgb}{0.24, 0.36, 0.64}
\definecolor{mysecondcolor}{rgb}{0.21, 0.64, 0.87}
\definecolor{myblue}{rgb}{.2,0.45,0.5} 
\definecolor{myorange}{rgb}{0.78,0.6,0.3}
\definecolor{mygreen}{rgb}{.2,0.38,0.16}
\definecolor{myalert}{rgb}{0.97,0.09,0.21}
\definecolor{myformulation}{rgb}{0.33, 0.29, 0.31}
\definecolor{myformulation_back}{rgb}{1, 0.97, 0.91}
\definecolor{hf}{rgb}{0.93, 0.57, 0.13} 
\definecolor{hf_2}{rgb}{1.0, 0.89, 0.77} 
\definecolor{hf_3}{rgb}{1.0, 0.22, 0.0} 
\definecolor{hf_4}{rgb}{1.0, 0.4, 0.1} 
\definecolor{burlywood}{rgb}{0.87, 0.72, 0.53}
\definecolor{burntorange}{rgb}{0.8, 0.33, 0.0}
\definecolor{burntsienna}{rgb}{0.91, 0.45, 0.32}
\definecolor{af}{rgb}{0.4, 0.53, 0.34}
\definecolor{af_2}{rgb}{0.74, 0.77, 0.47}
\definecolor{af_3}{rgb}{0.12, 0.3, 0.17}
\definecolor{af_4}{rgb}{0.03, 0.34, 0.25}
\definecolor{haf}{rgb}{0.6, 0.51, 0.48}
\definecolor{haf_2}{rgb}{1, 0.97, 0.91}
\definecolor{taf}{rgb}{0, 0.55, 0.5}
\definecolor{ajf}{rgb}{0.29, 0.59, 0.82}
\definecolor{hbf}{rgb}{0.87, 0.36, 0.51}
\definecolor{prussianblue}{rgb}{0.0, 0.19, 0.33}
\definecolor{regalia}{rgb}{0.32, 0.18, 0.5}
	
\definecolor{myred}{rgb}{0.7,0.15,0.15}
\definecolor{mygreen}{rgb}{0.13,0.55,0.13}
\definecolor{myblue}{rgb}{0.25,0.41,0.88}

\definecolor{vir_0}{rgb}{0.993248, 0.906157, 0.143936}
\definecolor{vir_1}{rgb}{0.565498, 0.84243 , 0.262877}
\definecolor{vir_2}{rgb}{0.20803 , 0.718701, 0.472873}
\definecolor{vir_3}{rgb}{0.128729, 0.563265, 0.551229}
\definecolor{vir_4}{rgb}{0.190631, 0.407061, 0.556089}
\definecolor{vir_5}{rgb}{0.267968, 0.223549, 0.512008}
\definecolor{vir_6}{rgb}{0.267004, 0.004874, 0.329415}

\definecolor{bw_6}{rgb}{0.05, 0.03, 0.53}
\definecolor{bw_5}{rgb}{0.42, 0.  , 0.66}
\definecolor{bw_4}{rgb}{0.69, 0.17, 0.56}
\definecolor{bw_3}{rgb}{0.88, 0.39, 0.38}
\definecolor{bw_2}{rgb}{0.99, 0.65, 0.21}
\definecolor{bw_1}{rgb}{0.94, 0.98, 0.13}

\begin{document}

\title{Reduced Order Hysteretic Magnetization Model for Composite Superconductors}

\author{Julien~Dular\footnote{Author to whom any correspondence should be addressed.}, Arjan~Verweij, Mariusz~Wozniak
}
\address{CERN, Geneva, Switzerland}
\ead{julien.dular@cern.ch}
\vspace{10pt}
\begin{indented}
\item[]\today
\end{indented}

\begin{abstract}
In this paper, we propose the Reduced Order Hysteretic Magnetization (ROHM) model to describe the magnetization and instantaneous power loss of composite superconductors subject to time-varying magnetic fields. Once the parameters of the ROHM model are fixed based on reference simulations, it allows to directly compute the macroscopic response of composite superconductors without the need to solve the detailed current density distribution. It can then be used as part of a homogenization method in large-scale superconducting models to significantly reduce the computational effort compared to detailed simulations. In this contribution, we focus on the case of a strand with twisted superconducting filaments subject to a time-varying transverse magnetic field. We propose two variations of the ROHM model: (i) a rate-independent model that reproduces hysteresis in the filaments, and (ii) a rate-dependent model that generalizes the first level by also reproducing dynamic effects due to coupling and eddy currents. We then describe the implementation and inclusion of the ROHM model in a finite element framework, discuss how to deduce the model parameters, and finally demonstrate the capabilities of the approach in terms of accuracy and efficiency over a wide range of excitation frequencies and amplitudes.
\end{abstract}
%




\vspace{2pc}
\noindent{\it Keywords: Reduced Order Method, Hysteresis Model, AC Loss, Magnetization.}


\ioptwocol



\section{Introduction}

Large-scale superconducting magnets are multi-scale structures made of multi-turn windings of superconducting cables, which themselves consist of multifilamentary strands or complex arrangements of tapes. The magnetic response of these systems is hysteretic and rate-dependent because of small-scale current density dynamics. Hysteresis is created by the irreversible motion of flux vortices among pinning centers in superconducting elements~\cite{bean1962magnetization, bean1964magnetization, brandt1992flux, richardson1994confirmation}. Rate-dependency is caused by eddy current and coupling current dynamics~\cite{campbell1982general, wilson1983superconducting, verweij1997electrodynamics}. 

Hysteresis, eddy current, and coupling current effects induce loss in transient regimes~\cite{campbell1982general} as well as field distortions~\cite{aleksa2004vector}. Their accurate description and computation is therefore crucial, e.g., for computing the load on the cryogenic system, temperature and stability margins, field errors, or for the design of quench protection devices.

Most approaches to describe the superconducting hysteresis are based on the definition of a relationship between electric field and current density, either non-smooth, as in the Bean model~\cite{bean1962magnetization}, or smooth, as in the power law model~\cite{rhyner1993magnetic}. The use of the power law model leads to very accurate results and versatile numerical models with, e.g., the finite element (FE) method~\cite{shen2020review}, which can also be combined with eddy current and coupling current models.

Besides these advantages, the power law introduces a strong nonlinearity in the equations, which makes numerical models of superconducting systems computationally demanding to solve, already for single strands or cables~\cite{lyly2011validation, zhao20173d, riva2023h, qiao20243d}. As a consequence, modelling large-scale superconducting systems in all their details down to the superconductor level is completely unrealistic in practice. It is often necessary to group conductors together in order to lower simulation time~\cite{zermeno20143d, vargas20223d}, and multi-scale procedures~\cite{denis2024ac} or domain decomposition methods~\cite{riva2023h, schnaubelt2024parallel} combined with parallel computing are being developed to speed up simulations.

Homogenization methods describing systems in terms of the average fields~\cite{marteau2023magnetic} without modelling the small-scale field distributions explicitly are also good candidates for obtaining even more efficient large-scale superconducting magnet models. The general idea of these methods consists in replacing complex conducting structures by fictitious, uniform, \textit{homogenized} materials with special properties. These properties are designed in order to reproduce directly the macroscopic effects of the small-scale current density distribution, e.g., magnetization and loss, without modelling that current density distribution explicitly. Once the homogenized material properties are defined by fitting their predictions with those of reference models, they can be used to reproduce accurately the response of large-scale systems with a strongly reduced computational cost compared to conventional detailed simulations.

Such homogenization methods have been extensively studied in non-superconducting electromagnetic systems, such as periodical structures with bundles of wires of any shape~\cite{el1997homogenization, meunier2010homogenization, gyselinck2006nonlinear}, described in the frequency or time domain~\cite{gyselinck2005frequency, sabariego2008time}. They rely on the definition of equivalent permeability and conductivity values (or functions), to account for proximity effect and skin effect, respectively. 

Following a similar approach for superconducting systems requires to define history-dependent material properties in order to reproduce a hysteretic response. This is the objective of this paper.


A few hysteresis models have been proposed for modelling superconductors. A vector model is proposed in~\cite{aleksa2004vector, vollinger2002superconductor} for field quality computation in accelerator magnets. This model describes the magnetization of superconducting filaments using nested magnetization ellipses. Another hysteresis model based on the Preisach model~\cite{preisach1935magnetische} was proposed in~\cite{sjostrom2001hysteresis, sjostrom2003equivalent} in the form of an equivalent circuit element. Yet another vector model based on a variational approach was described in~\cite{badia2002vector}. These models however only describe rate-independent hysteresis and are therefore not suited for modelling transient effects in composite superconductors.

The literature on hysteresis models for ferromagnetic materials is much more extensive. Among the most popular models are the Jiles-Atherton~\cite{jiles1986theory}, Preisach~\cite{preisach1935magnetische} models, and rate-dependent models such as the Chua-Stromsmoe model~\cite{chua1970lumped, chua1972generalized}. In 1997, Bergqvist proposed a thermodynamically consistent approach to model hysteresis~\cite{bergqvist1997magnetic}, which led to the development of the energy-based model~\cite{henrotte2006energy, henrotte2006dynamical, jacques2018energy}.

The energy-based model offers several advantages compared to other models. First, it gives a direct access to instantaneous dissipated and stored power by conveniently separating the magnetic field into several contributions. Second, it offers a high flexibility and modularity thanks to an approach involving several sub-elements that are combined to form the final model. Third, it is consistently defined as a vector model from the beginning~\cite{henrotte2006energy}, and hence applies to two-dimensional (2D) or three-dimensional (3D) settings. Fourth, it can be directly generalized to include rate-dependent effects~\cite{henrotte2006dynamical}. Finally, the equations of this model consist in simple explicit tests, which leads to straightforward implementations and very efficient simulations.

In this paper, we adapt and extend the state-of-the-art energy-based model to composite superconductor hysteresis, paying attention to preserving all the above-mentioned advantages. In this first contribution, we focus on the case of a superconducting multifilamentary strand subject to a transverse magnetic field, i.e., a field perpendicular to the strand axis, which is relevant for most magnet applications. The multifilamentary strand is also a challenging problem. Indeed, in addition to the filament hysteresis effect, it also involves two types of rate-dependent effects, interfilament coupling currents and eddy currents in the strand matrix, as well as filament hysteresis change due to coupling and eddy currents.

In this paper, we assume that the strand carries no transport current. Accounting for transport current and the computing the associated voltage is necessary for the homogenization of a complete magnet winding, but will be the focus of a further work. We also consider a constant temperature and do not model thermal effects.

To describe the response of a multifilamentary strand to an external magnetic field, we propose the Reduced Order Hysteretic Magnetization (ROHM) model. We present two variations of it: (i) a rate-independent model which describes the superconducting hysteresis, and (ii) a rate-dependent model, which generalizes the former by also describing eddy current and coupling current effects, in addition to the superconducting hysteresis. The ROHM model consists in describing the strand as a homogenized \textit{non-conducting} magnetic material described by a hysteretic constitutive law $\b = \bhyst(\h)$ between the magnetic field $\h$ and the magnetic flux density $\b$, which directly describes the magnetization and instantaneous power loss in the strand. By construction, the ROHM model can be easily implemented in any FE framework that allows for user-defined material properties.

Although we apply it on multifilamentary strands only, the ROHM model can be adapted to a variety of superconductors or composite superconductors, such as strands, tapes, stacks of tapes, cables, or bulks.

This paper is structured as follows. In Section~\ref{sec_context}, we describe the response of a multifilamentary strand in terms of magnetization and loss for transverse fields with a range of frequencies and amplitudes, in order to motivate the design of the ROHM model. In Section~\ref{sec_cells}, we present the equations defining the rate-independent and rate-dependent variations of the ROHM model. In Section~\ref{sec_resolution}, we discuss the implementation of these equations and we describe how to include them in a 2D FE framework with a $\phi$\nobreakdash-formulation. We then propose in Section~\ref{sec_paramIdentification} a parameter identification approach for fixing the ROHM model parameters based on reference solutions. Finally, in Sections~\ref{sec_application_static} and \ref{sec_multiStrand}, we apply the ROHM model in rate-independent and rate-dependent situations, respectively, and demonstrate its potential as a homogenized model.

\section{Multifilamentary Strand Dynamics}\label{sec_context}

In this section, we describe the response of a composite superconducting strand subject to a transverse magnetic field in terms of the induced power loss and magnetization. To this end, we model the detailed current density distribution inside the strand with a FE method, the CATI method~\cite{dular2024coupled} implemented in FiQuS~\cite{vitrano2023open, getdp, gmsh}, and we analyze the different dynamics resulting from the strand composite structure.



\subsection{Problem definition}\label{sec_problemDefinition}

\begin{figure}[h!]
\begin{center}
\includegraphics[width=\linewidth]{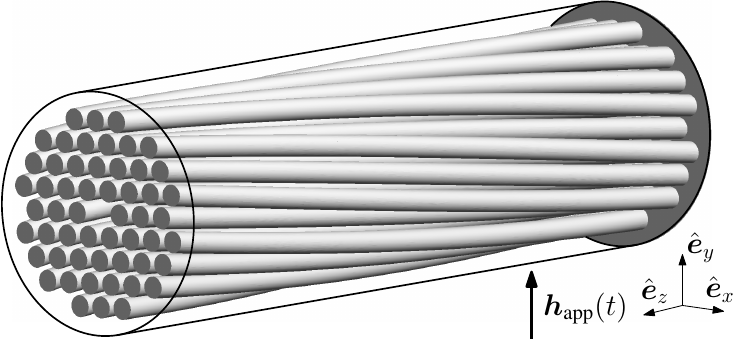}
\caption{Composite strand with twisted superconducting filaments embedded in a conducting matrix (cylindrical outline) subject to a time-varying transverse magnetic field $\hs(t)$.}
\label{strand_3D}
\end{center}
\end{figure}

We consider a multifilamentary strand such as in Fig.~\ref{strand_3D}. It consists of twisted superconducting filaments embedded in a cylindrical conducting matrix. The strand carries no transport current but is subject to a time-varying transverse magnetic field $\hs(t) = \hmax \sin(2\pi f t)\ey$ of amplitude $\hmax$ (A/m) and frequency $f$ (Hz).

\subsection{Magnetization and power loss}

We focus on magnetization and power loss. These are the two quantities to be reproduced with the ROHM model.

The average magnetization vector $\vec m$ (A/m) is defined as the magnetic dipole moment per unit volume~\cite{griffiths2023introduction}. Based on the current density distribution on a given cross section, we have
\begin{align}\label{eq_magnetization}
\vec m = 2\cdot \frac{1}{2a}\int_{S} \vec x \times \j\, \text{d}S,
\end{align}
with $a$ the surface area of the strand cross section $S$, $\vec x$ the position vector, and $\j$ the current density. The multiplication by $2$ accounts for current loops closing at infinity, as justified in~\ref{app_magnetization}.

The instantaneous power loss per unit length $P_\text{tot}$ (W/m) is obtained by integrating the instantaneous power loss density, or Joule loss density, over the strand cross section, including the filaments and the matrix,
\begin{align}\label{eq_ptot}
P_\text{tot} = \int_{S} \j\cdot \e\, \text{d}S,
\end{align}
with $\e =\rho \j$ the electric field and $\rho$ the resistivity. The energy loss per cycle and per unit length (J/m) is evaluated as
\begin{align}
Q_\text{tot} = \int_{1/f}^{2/f} P_\text{tot}(t)\, \text{d}t.
\end{align}
To make the interpretation of the loss easier, and to aid the construction of the ROHM model, the total energy loss per cycle $Q_\text{tot}$ is decomposed in three contributions: hysteresis, coupling, and eddy loss. The hysteresis loss is the loss induced by currents flowing in the superconducting filaments. The coupling loss is the loss induced by currents flowing in the conducting matrix perpendicular to $\ez$ (the coupling currents). Finally, the eddy loss is the loss induced by currents flowing in the conducting matrix along $\ez$.

Magnetization and loss are related to each other. Indeed, the area inside a closed magnetization loop is proportional to the total loss per cycle~\cite{bossavit2000remarks,goldfarb1986internal}. We have, in terms of the applied magnetic field $\hs$,
\begin{align}\label{eq_lossPerCycle_detailed}
Q_\text{tot} = a \oint \mu_0 \hs \cdot \text{d}\vec m. 
\end{align}

\subsection{Dynamic response of the strand}\label{sec_dynamicResponse}

We now describe the strand response as a function of the field amplitude and frequency, and highlight its hysteretic behavior as well as the associated rate-dependency.

For illustration, we consider a strand made of $54$ Nb-Ti filaments (diameter $90$~$\upmu$m) twisted with a twist pitch length of $19$~mm and embedded in a Cu matrix with a diameter of $1$~mm (same geometry as in~\cite{dular2024coupled}). We fix the temperature to $1.9$~K. The resistivity of the Nb-Ti filaments is described by the power law~\cite{rhyner1993magnetic} with power index $n=30$ and a field-dependent critical current density $\jc(\b)$~\cite{zachou2024}, with $\b$ (T) the local magnetic flux density, defined as $\b = \mu_0 \h$, with $\mu_0 = 4\pi\times 10^{-7}$~H/m. The resistivity of the copper matrix accounts for magneto-resistance, with a residual resistivity ratio $\text{RRR} = 100$ at zero field~\cite{zachou2024}.

\subsubsection{Rate-independent regime}

\begin{figure}[h!]
    \centering
\centering
\tikzsetnextfilename{magnCurves_2T_uncoupled}
\begin{tikzpicture}[trim axis left, trim axis right][font=\small]
\pgfplotsset{set layers}
 \begin{axis}[
tick label style={/pgf/number format/fixed},
width=\linewidth,
height=5cm,
grid = major,
grid style = dotted,
xmin=-2.1, 
xmax=2.1,
ytick={-0.1, -0.05, 0, 0.05, 0.1},
yticklabels={$-0.2$, $-0.1$, $0$, $0.1$, $0.2$},
xlabel={Applied magnetic field $\mu_0\hsS$ (T)},
ylabel={Magnetization $\mu_0 m$ (T)},
ylabel style={yshift=-2.8em},
xlabel style={yshift=0.5em},
xticklabel style={yshift=0.1em},
yticklabel style={xshift=0em},
yticklabel pos=right,
legend columns=3,
transpose legend,
legend style={at={(0.01, 0.051)}, cells={anchor=west}, anchor=south west, draw=none,fill opacity=0, text opacity = 1}
]
\addplot[vir_6, thick] 
table[x=b,y=magn_tot]{data/Magnetization_f0.01_b2.txt};
\legend{$f=0.01$ Hz}
\end{axis}
\end{tikzpicture}
\vspace{-0.2cm}
\caption{Strand magnetization for $\mu_0\hmax = 2$~T, $f = 0.01$~Hz.}
    \label{magnCurves_2T_uncoupled}
\end{figure}
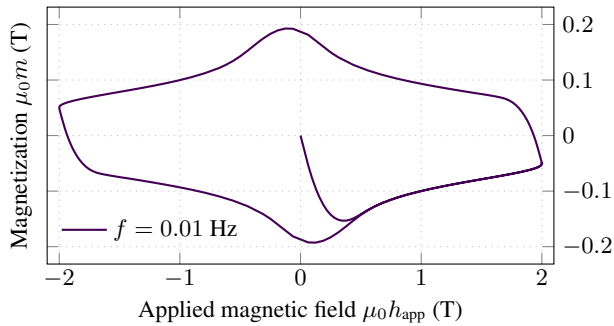

At low frequencies (here for $f\lesssim 0.01$~Hz), the rate of change of the transverse magnetic field generates negligible coupling and eddy currents in the matrix~\cite{campbell1982general, denis2024ac}. In this regime, the superconducting filaments behave almost as uncoupled and most of the loss is due to hysteresis in them. A typical magnetization loop is shown in Fig.~\ref{magnCurves_2T_uncoupled}. The magnetization decreases when the field increases as a result of the field-dependent critical current density.

Strictly speaking, the finite value of $n$ creates a small rate-dependency, but in good approximation, the response can be considered rate-independent. A rate-independent ROHM model can therefore be used to describe the strand response in this regime.

\subsubsection{Rate-dependent regime}

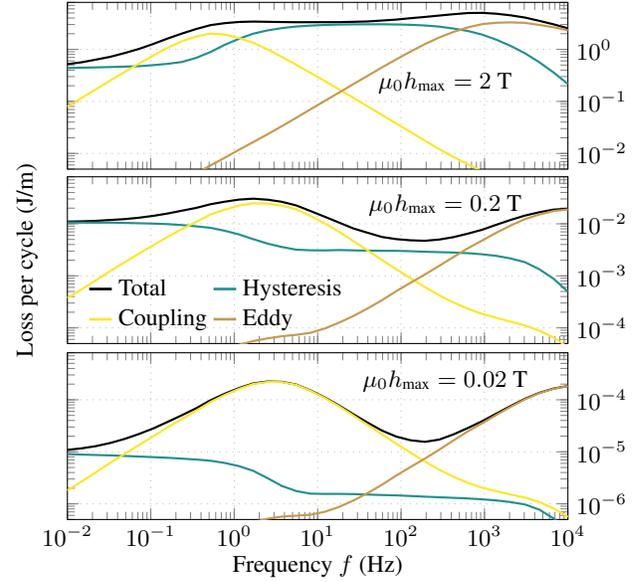
\begin{figure}[h!]
\centering
 \begin{subfigure}[b]{0.99\linewidth}  
 \centering
\tikzsetnextfilename{strand_lossCycle_2p0T}
\begin{tikzpicture}[trim axis left, trim axis right][font=\small]
\pgfplotsset{set layers}
 	\begin{loglogaxis}[
	tick label style={/pgf/number format/fixed},
    width=\linewidth,
    height=3.8cm,
    grid = major,
    grid style = dotted,
    ymin=5e-3, 
    ymax=8,
    xmin=0.01, 
    xmax=10000,
    xticklabels={},
    ylabel style={yshift=-2.8em},
    xlabel style={yshift=0.5em},
    xticklabel style={yshift=0.1em},
    yticklabel style={xshift=0em},
    yticklabel pos=right,
    legend columns=2,
    legend style={at={(0.38, 0.02)}, cells={anchor=west}, anchor=south, draw=none,fill opacity=0, text opacity = 1}
    ]
    \addplot[black, thick] 
    table[x=f,y=total]{data/strand_lossCycle_2p0T.txt};
        \addplot[vir_3, thick] 
    table[x=f,y=fil]{data/strand_lossCycle_2p0T.txt};
        \addplot[vir_0, thick] 
    table[x=f,y=coupling]{data/strand_lossCycle_2p0T.txt};
        \addplot[myorange, thick] 
    table[x=f,y=eddy]{data/strand_lossCycle_2p0T.txt};
\node[anchor=south] at (axis cs: 337, 0.1) {$\mu_0\hmax = 2$ T};
    \end{loglogaxis}
\end{tikzpicture}%
\end{subfigure}
        \hfill\vspace{-0.4cm}
 \begin{subfigure}[b]{0.99\linewidth}  
 \centering
\tikzsetnextfilename{strand_lossCycle_0p2T}
\begin{tikzpicture}[trim axis left, trim axis right][font=\small]
\pgfplotsset{set layers}
 	\begin{loglogaxis}[
	tick label style={/pgf/number format/fixed},
    width=\linewidth,
    height=3.8cm,
    grid = major,
    grid style = dotted,
    ymin=5e-5, 
    ymax=8e-2,
    xmin=0.01, 
    xmax=10000,
    xticklabels={},
    ylabel={Loss per cycle (J/m)},
    ylabel style={yshift=-2.2em},
    xlabel style={yshift=0.5em},
    xticklabel style={yshift=0.1em},
    yticklabel style={xshift=0em},
    yticklabel pos=right,
    legend columns=2,
    legend style={at={(0.3, 0.02)}, cells={anchor=west}, anchor=south, draw=none,fill opacity=0, text opacity = 1, legend image code/.code={\draw[##1,line width=1pt] plot coordinates {(0cm,0cm) (0.3cm,0cm)};}}
    ]
    \addplot[black, thick] 
    table[x=f,y=total]{data/strand_lossCycle_0p2T.txt};
        \addplot[vir_3, thick] 
    table[x=f,y=fil]{data/strand_lossCycle_0p2T.txt};
        \addplot[vir_0, thick] 
    table[x=f,y=coupling]{data/strand_lossCycle_0p2T.txt};
        \addplot[myorange, thick] 
    table[x=f,y=eddy]{data/strand_lossCycle_0p2T.txt};
    \legend{Total, Hysteresis, Coupling, Eddy}
    \node[anchor=south] at (axis cs: 337, 0.01) {$\mu_0\hmax = 0.2$ T};
    \end{loglogaxis}
\end{tikzpicture}%
\end{subfigure}
        \hfill\vspace{-0.2cm}
 \begin{subfigure}[b]{0.99\linewidth}  
        \centering
\tikzsetnextfilename{strand_lossCycle_0p02T}
\begin{tikzpicture}[trim axis left, trim axis right][font=\small]
\pgfplotsset{set layers}
 	\begin{loglogaxis}[
	tick label style={/pgf/number format/fixed},
    width=\linewidth,
    height=3.8cm,
    grid = major,
    grid style = dotted,
    ymin=5e-7, 
    ymax=8e-4,
    xmin=0.01, 
    xmax=10000,
	xlabel={Frequency $f$ (Hz)},
    ylabel style={yshift=-2.8em},
    xlabel style={yshift=0.5em},
    xticklabel style={yshift=0.1em},
    yticklabel style={xshift=0em},
    yticklabel pos=right,
    legend columns=2,
    legend style={at={(0.38, 0.03)}, cells={anchor=west}, anchor=south, draw=none,fill opacity=0, text opacity = 1, legend image code/.code={\draw[##1,line width=1pt] plot coordinates {(0cm,0cm) (0.3cm,0cm)};}}
    ]
    \addplot[black, thick] 
    table[x=f,y=total]{data/strand_lossCycle_0p02T.txt};
        \addplot[vir_3, thick] 
    table[x=f,y=fil]{data/strand_lossCycle_0p02T.txt};
        \addplot[vir_0, thick] 
    table[x=f,y=coupling]{data/strand_lossCycle_0p02T.txt};
        \addplot[myorange, thick] 
    table[x=f,y=eddy]{data/strand_lossCycle_0p02T.txt};
\node[anchor=south] at (axis cs: 337, 0.0001) {$\mu_0\hmax = 0.02$ T};
    \end{loglogaxis}
\end{tikzpicture}%
\end{subfigure}
 \hfill
\vspace{-0.2cm}
\caption{Total loss per cycle and its contributions as a function of the frequency $f$, for three different magnetic field amplitudes. The legend is the same for the three subfigures. Results from~\cite{dular2024coupled}.}
        \label{strand_lossCycle}
\end{figure}
\begin{figure}[h!]
    \centering
\centering
\tikzsetnextfilename{magnCurves_2T}
\begin{tikzpicture}[trim axis left, trim axis right][font=\small]
\pgfplotsset{set layers}
 \begin{axis}[
tick label style={/pgf/number format/fixed},
width=\linewidth,
height=7.5cm,
grid = major,
grid style = dotted,
ymin=-4, 
ymax=4,
xmin=-2.1, 
xmax=2.1,
ytick={-4, -3, -2, -1, 0, 1, 2, 3, 4},
yticklabels={$-4$, $-3$, $-2$, $-1$, $0$, $1$, $2$, $3$, $4$},
xlabel={Applied magnetic field $\mu_0\hsS$ (T)},
ylabel={Magnetization $\mu_0 m$ (T)},
ylabel style={yshift=-2.8em},
xlabel style={yshift=0.5em},
xticklabel style={yshift=0.1em},
yticklabel style={xshift=0em},
yticklabel pos=right,
legend columns=4,
transpose legend,
legend style={at={(0.01, 0.01)}, cells={anchor=west}, anchor=south west, draw=none,fill opacity=0, text opacity = 1}
]
\addplot[vir_6, thick] 
table[x=b,y=magn_tot]{data/Magnetization_f0.01_b2_hin.txt};
    \addplot[vir_5, thick] 
table[x=b,y=magn_tot]{data/Magnetization_f0.1_b2_hin.txt};
    \addplot[vir_4, thick] 
table[x=b,y=magn_tot]{data/Magnetization_f1_b2_hin.txt};
    \addplot[vir_3, thick] 
table[x=b,y=magn_tot]{data/Magnetization_f10_b2_hin.txt};
\addlegendimage{empty legend}
\addlegendentry{} 
    \addplot[vir_2, thick] 
table[x=b,y=magn_tot]{data/Magnetization_f100_b2_hin.txt};
    \addplot[vir_1, thick] 
table[x=b,y=magn_tot]{data/Magnetization_f1000_b2_hin.txt};
    \addplot[vir_0, thick] 
table[x=b,y=magn_tot]{data/Magnetization_f10000_b2_hin.txt};

\legend{$f=0.01$ Hz, $f=0.1$ Hz, $f=1$ Hz, $f=10$ Hz, \empty, $f=10^{2}$ Hz, $f=10^{3}$ Hz, $f=10^{4}$ Hz}
\end{axis}
\end{tikzpicture}
\vspace{-0.2cm}
\caption{Strand magnetization for $\mu_0\hmax = 2$~T and for different frequencies $f$. Results from~\cite{dular2024coupled} using Eq.~\eqref{eq_magnetization}. The curve for $f= 0.01$~Hz is shown in more detail in Fig.~\ref{magnCurves_2T_uncoupled}.}
    \label{magnCurves_2T}
\end{figure}
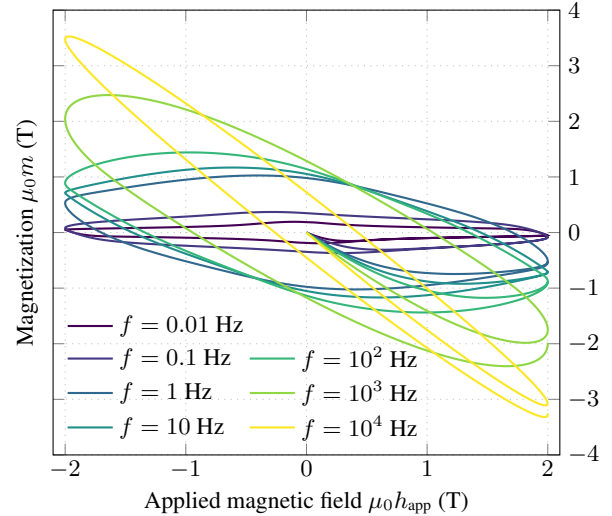

When the frequency increases, the response of the strand is no longer rate-independent. As shown in Fig.~\ref{strand_lossCycle}, coupling and eddy loss can no longer be neglected, and the hysteresis loss varies significantly with frequency. As a result, the total loss per cycle is strongly rate-dependent. The magnetization also strongly depends on the frequency, as shown in Fig.~\ref{magnCurves_2T}, where the $y$-component of the magnetization vector, $m$, is shown as a function of $\mu_0\hsS(t)$ for different frequencies. 

To reproduce magnetization and loss in this regime, there is a clear need for a rate-dependent ROHM model. The design of such a model benefits from a good understanding of the physics behind the curves in Figs.~\ref{strand_lossCycle} and \ref{magnCurves_2T}, which we briefly summarize below.

For low field amplitudes, the dynamics of the coupling currents and of the associated loss is well described by analytical models~\cite{campbell1982general, morgan1970theoretical}. The coupling loss follows a typical bell curve with a maximum at a frequency $f_\text{c}$.

With increasing field amplitude, the peak of the coupling loss curve shifts towards lower frequencies, as can be seen in Fig.~\ref{strand_lossCycle} for $\mu_0\hmax = 2$~T. This is associated with the saturation of superconducting filaments, which limits the coupling currents amplitude, and to the loss of the diamagnetic effect in saturated filaments, as discussed in~\cite{campbell1982general}.

The coupling current dynamics influences the filament magnetization and hysteresis loss. At low frequencies, $f\lesssim f_\text{c}$, filaments are mostly uncoupled, whereas at higher frequencies, $f\gtrsim f_\text{c}$, they exchange currents via coupling currents. As a result, the overall magnetization increases, see for example the situation for $f=10$~Hz in Fig.~\ref{magnCurves_2T}, and the hysteresis loss is affected. Whether the hysteresis loss per cycle for coupled filaments is higher or lower than that for uncoupled filaments depends on the applied field amplitude, as can be seen in Fig.~\ref{strand_lossCycle}.

At even higher frequencies, eddy currents in the conducting matrix become the dominating factor for loss and magnetization. As a result of eddy currents and the skin effect, the inner part of the strand, containing the filaments, is shielded from the external field and the hysteresis loss decreases accordingly. In that regime, the magnetization curve approaches the shape of an ellipse~\cite{bossavit2000remarks} that becomes thinner and thinner with increasing frequency, and its slope approaches $-2$ because of demagnetization effects~\cite{goldfarb1986internal}.

\section{Reduced Order Hysteretic Magnetization Model}\label{sec_cells}

The simulation time for each of the simulations presented in the previous section is of the order of one hour with the CATI method~\cite{dular2024coupled}. This is fast and convenient enough to get a good understanding of a single strand response with a detailed description of the current density distribution (as discussed in~\cite{dular2024coupled} or shown later in Fig.~\ref{homogenization_in_FE}). However, to simulate transients in a full-scale magnet cross section containing thousands of strands, one cannot afford models involving such a detailed description at the strand level.

Instead, one can use an alternative model that produces accurate macroscopic response in terms of power loss and magnetization, which are both history-dependent and rate-dependent, but that does not rely on a detailed calculation of the current density. We present such a model in this section: the Reduced Order Hysteretic Magnetization (ROHM) model. This model is inspired from the energy-based model for ferromagnetic materials~\cite{henrotte2006energy, henrotte2006dynamical}. 

We start in Section~\ref{sec_homogenizationConcept} by clarifying how the ROHM model can be used to model a superconducting strand.

We then define the building blocks, referred to as \textit{cells}, that will be used for the construction of the ROHM model. Each cell defines a local and elementary hysteretic relationship between the magnetic field $\h$ as an input, and the magnetic flux density $\b$ as an output. We present in Section~\ref{sec_staticHystCell} a rate-independent cell, that we call the superconductor cell (\SC). In Section~\ref{sec_cell_coupling}, we add components to account for eddy current and coupling current effects, resulting in a rate-dependent cell, that we call the composite superconductor cell (\CSC). Inside each cell, the magnetic field vector is expressed as the sum of distinct components: 2 components for the \SC, and 4 components for the \CSC. These components describe the interplay between different magnetization mechanisms and allow for a clear separation between stored and dissipated energies.

A single cell is not sufficient for an accurate description of the strand response. We show in Section~\ref{sec_multiCell} how several cells can be combined into a \textit{chain of cells}, defining the complete ROHM model.


\subsection{Homogenized model and internal field}\label{sec_homogenizationConcept}

\begin{figure}[h!]
\begin{center}
\includegraphics[width=\linewidth]{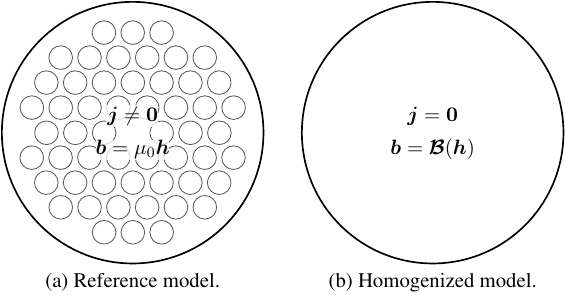}
\caption{Homogenization concept without transport current.} 
\label{homogenizationConcept}
\end{center}
\end{figure}

The idea of the ROHM model is to replace the detailed cross section of the strand by a plain, homogenized material, as illustrated in Fig.~\ref{homogenizationConcept}. The homogenized material is assumed to be non-conducting, but magnetic. It is described by a constitutive law $\b = \bhyst(\h)$ between the local magnetic field $\h$ and the local magnetic flux density $\b$. The constitutive law is carefully designed such that it produces a magnetization vector and power loss value that are, on average, equivalent to those obtained with a detailed model~\cite{bossavit2000remarks}.

In the homogenized model, the magnetization vector $\m$ is no longer computed with the integral of Eq.~\eqref{eq_magnetization} as there is no current density in the material. Instead, with fields $\b$ and $\h$ that satisfy $\b = \bhyst(\h)$, the magnetization vector $\m$ (A/m) is defined by~\cite{bossavit2000remarks}
\begin{align}
\b = \mu_0 \h + \mu_0 \m.
\end{align}
Similarly, the power loss is no longer computed as a Joule loss as in Eq.~\eqref{eq_ptot}, but is now linked with the magnetic power density $p_\text{tot} = \h \cdot \dot \b$. Part of this power is related to stored magnetic energy, and hence is not associated with loss. The complementary part is related to dissipated energy. One major advantage of the energy-based hysteresis model used in this work is that it gives access to both parts (stored and dissipated) separately at all time~\cite{henrotte2006energy}.

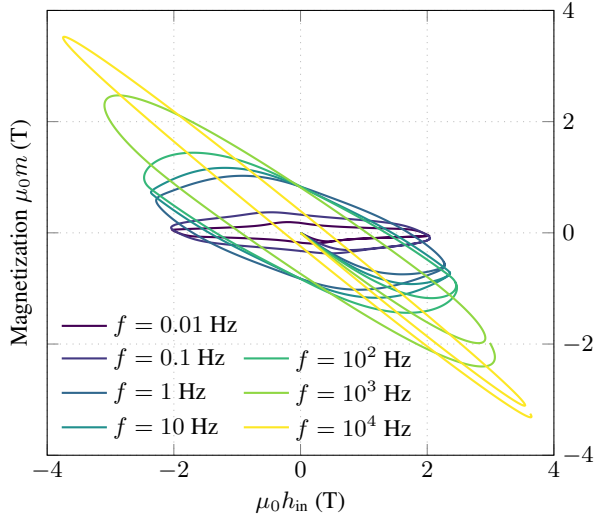
\begin{figure}[h!]
    \centering
\centering
\tikzsetnextfilename{magnCurves_2T_hin}
\begin{tikzpicture}[trim axis left, trim axis right][font=\small]
\pgfplotsset{set layers}
 \begin{axis}[
tick label style={/pgf/number format/fixed},
width=\linewidth,
height=7.5cm,
grid = major,
grid style = dotted,
ymin=-4, 
ymax=4,
xmin=-4, 
xmax=4,
xlabel={$\mu_0\hinS$ (T)},
ylabel={Magnetization $\mu_0 m$ (T)},
ylabel style={yshift=-2.8em},
xlabel style={yshift=0.5em},
xticklabel style={yshift=0.1em},
yticklabel style={xshift=0em},
yticklabel pos=right,
legend columns=4,
transpose legend,
legend style={at={(0.01, 0.011)}, cells={anchor=west}, anchor=south west, draw=none,fill opacity=0, text opacity = 1}
]
\addplot[vir_6, thick] 
table[x=mu0hin,y=magn_tot]{data/Magnetization_f0.01_b2_hin.txt};
    \addplot[vir_5, thick] 
table[x=mu0hin,y=magn_tot]{data/Magnetization_f0.1_b2_hin.txt};
    \addplot[vir_4, thick] 
table[x=mu0hin,y=magn_tot]{data/Magnetization_f1_b2_hin.txt};
    \addplot[vir_3, thick] 
table[x=mu0hin,y=magn_tot]{data/Magnetization_f10_b2_hin.txt};
\addlegendimage{empty legend}
\addlegendentry{} 
    \addplot[vir_2, thick] 
table[x=mu0hin,y=magn_tot]{data/Magnetization_f100_b2_hin.txt};
    \addplot[vir_1, thick] 
table[x=mu0hin,y=magn_tot]{data/Magnetization_f1000_b2_hin.txt};
    \addplot[vir_0, thick] 
table[x=mu0hin,y=magn_tot]{data/Magnetization_f10000_b2_hin.txt};

\legend{$f=0.01$ Hz, $f=0.1$ Hz, $f=1$ Hz, $f=10$ Hz, \empty, $f=10^{2}$ Hz, $f=10^{3}$ Hz, $f=10^{4}$ Hz}
\end{axis}
\end{tikzpicture}
\vspace{-0.2cm}
\caption{Strand magnetization for $\mu_0\hmax = 2$~T. The results are the same as those in Fig.~\ref{magnCurves_2T} but in terms of $\mu_0\hinS$ instead of $\mu_0\hsS$ (see Eq.~\eqref{eq_hin_definition}).}
    \label{magnCurves_2T_hin}
\end{figure}

The relationship $\b = \boldsymbol{\mathcal{B}}(\h)$ is written in terms of the local fields $\h$ and $\b$, internal to the strand. Due to the demagnetization effect, the internal magnetic field $\h$ is in general not equal to the applied field $\hs$. In the case of a round strand subject to a transverse external field, the magnetization vector $\m$ is uniform in the strand and the demagnetization factor is equal to $1/2$, such that we have~\cite{goldfarb1986internal}
\begin{align}
\h = \hs - \frac{1}{2} \vec m.
\end{align}
For this reason, the magnetization curves produced by the ROHM model $\b = \boldsymbol{\mathcal{B}}(\h)$ are not to be compared with those depicted in Figs.~\ref{magnCurves_2T_uncoupled} and \ref{magnCurves_2T}, but rather with different ones, drawn as a function of 
\begin{align}\label{eq_hin_definition}
\hin = \hs - \frac{1}{2}\vec m,
\end{align}
where $\hin$ has no immediate meaning in the detailed strand model per se, but is the local magnetic field that the ROHM model will see inside the strand. These curves are shown in Fig.~\ref{magnCurves_2T_hin}. At high frequencies ($f\gtrsim 10^4$~Hz), they approach the shape of an ellipse with a slope of $-1$. The area of the closed loops is preserved between Fig.~\ref{magnCurves_2T} and Fig.~\ref{magnCurves_2T_hin}, as shown in~\ref{app_app_and_in_fields}.

\subsection{Superconductor cell - \SC}\label{sec_staticHystCell}

\begin{figure}[h!]
      \begin{subfigure}[b]{0.49\linewidth}  
            \centering 
		\includegraphics[width=\textwidth]{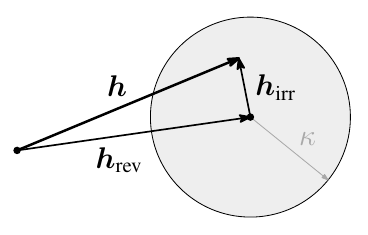}
		\caption{$\|\h - \hrev\| < \ku$.}
		\label{hystModel_static_inside}	
      \end{subfigure}
      \begin{subfigure}[b]{0.49\linewidth}  
            \centering 
		\includegraphics[width=\textwidth]{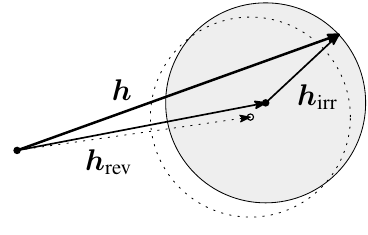}
		\caption{$\|\h - \hrev\| = \ku$.}
		\label{hystModel_static_edge}	
      \end{subfigure}
\caption{\SC, adapted from~\cite{henrotte2006energy, jacques2018energy}. Illustration in 2D of the magnetic field decomposition into reversible and irreversible fields. The dotted elements in (b) refer to the situation at a previous time.}
\label{hystModel_static}
\end{figure}

As a first step, we present a rate-independent hysteresis cell~\cite{henrotte2006energy}: the superconductor cell, or \SC. In this cell, the magnetic field $\h$ is decomposed into a reversible field, $\hrev$, and an irreversible field\footnote{The irreversible field $\hirr$ has nothing to do with the irreversibility field of a superconductor.}, $\hirru$:
\begin{align}\label{eq_staticCellHDecomp}
\h = \hrev + \hirru.
\end{align}
The reversible field defines the magnetic flux density
\begin{align}\label{eq_bhrev}
\b = \mu_0 \hrev.
\end{align}
The irreversible field creates hysteresis by introducing history dependence. Its amplitude $\|\hirru\|$ is bounded by a value $\ku$ (A/m), the \textit{irreversibility parameter}. From a known magnetic field $\h$, and a given reversible field $\hrev$ that has been established previously (as a function of the history of $\h$), the irreversible field $\hirru$ is determined as follows:
\begin{align}\label{eq_constitutive_static}
\hirru &= \left\{\begin{aligned}
&\h - \hrev,\quad &&\text{if } \|\h - \hrev\| < \ku,\\
&\ku \ \dot \b/\|\dot \b\|,\quad &&\text{if } \|\h - \hrev\| = \ku,\\
\end{aligned}\right.
\end{align}
where the dot notation (as in $\dot \b$) represents the time derivative. An illustration of this equation is given in Fig.~\ref{hystModel_static}. While the driving field $\h$ evolves inside the sphere of radius $\ku$ centered at $\hrev$, the reversible field $\hrev$ stays constant (and so does $\b = \mu_0 \hrev$), and the irreversible field is defined accordingly, see Fig.~\ref{hystModel_static}(a). If $\h$ reaches the surface of the sphere, then, the reversible field $\hrev$ (and, hence, also $\b$) must evolve in order to maintain the condition $\|\hirru\|\le \ku$ valid, and $\hirru$ is established parallel to $\dot \b$, see Fig.~\ref{hystModel_static}(b).

The total power density $\ptot$ (W/m$^3$) is expressed as
\begin{align}
\ptot &= \h \cdot \dot \b\notag\\
&= \underbrace{\hrev \cdot \dot \b}_{\prev} + \underbrace{\hirru \cdot \dot \b}_{\pirru}.
\end{align}
The reversible power $\prev$ is the time derivative of the stored magnetic energy density,
\begin{align}\label{eq_prev}
\prev = \hrev \cdot  \mu_0\dhrev = \derd{}{t}\paren{\frac{1}{2} \hrev\cdot\mu_0 \hrev},
\end{align}
while the irreversible power $\pirr$ is always non-negative and is associated with the dissipated energy density,
\begin{align}\label{eq_pirr}
\pirru = \hirru \cdot \dot \b = \ku \|\dot\b\| = \ku \|\mu_0 \dhrev\|.
\end{align}
It is one of the main benefits of the energy-based approach that the dissipated energy is clearly separated from the stored energy~\cite{henrotte2006energy}. 

\begin{figure}[h!]
\begin{center}
\includegraphics[width=0.4\linewidth]{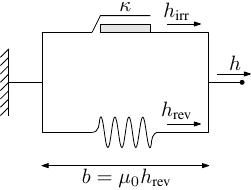}
\caption{Mechanical analogy~\cite{henrotte2006energy} of the \SC: a parallel connection of a dry friction element and a restoring spring. In the analogy, the system is subject to a force $h$ which causes an elongation $b=\mu_0 h_\text{rev}$.}
\label{mechanical_analogy_static}
\end{center}
\end{figure}

A mechanical analogy of the \SC\ in a one-dimensional setting is shown in Fig.~\ref{mechanical_analogy_static}. It represents a mechanical system made of a dry friction element in parallel with a restoring spring~\cite{henrotte2006energy}.

In practice, the magnetization of superconductors decreases with increasing field as a consequence of the decreasing critical current density $\jc$, as is shown in Fig.~\ref{magnCurves_2T_uncoupled}. In order to reproduce this effect with the \SC, one can make the irreversibility parameter $\kappa$ a function of the magnetic flux density $\b$, by defining
\begin{align}\label{eq_jc_scaling}
\ku(\b) = f_\kappa(\b) \bar\kappa,
\end{align}
with $\bar\kappa$ (A/m) a constant and $f_\kappa$ a scaling function, smoothly evolving from $1$ at zero field to $0$ at large fields. The expression of $f_\kappa$ is function of the geometry and must be determined at the parameter identification step, as discussed in Section~\ref{sec_paramIdentification}. The introduction of a field-dependent irreversibility parameter brings an additional nonlinearity in the equations, which has to be addressed during the numerical simulation.

\begin{figure}[h!]
    \centering
\begin{subfigure}[b]{0.99\linewidth}
\centering
\tikzsetnextfilename{cell_static_f1_b2_b}
\begin{tikzpicture}[trim axis left, trim axis right][font=\small]
\pgfplotsset{set layers}
 \begin{axis}[
tick label style={/pgf/number format/fixed},
width=0.9\textwidth,
height=6.3cm,
grid = major,
grid style = dotted,
ymin=-2, 
ymax=2,
xmin=-2.2, 
xmax=2.2,
ylabel={$b$ (T)},
ylabel style={yshift=-2em},
xlabel style={yshift=0.5em},
xticklabel style={yshift=0.1em},
yticklabel style={xshift=0em},
legend style={at={(0.01, 0.99)}, cells={anchor=west}, anchor=north west, draw=none,fill opacity=0, text opacity = 1, legend image code/.code={\draw[##1,line width=1pt] plot coordinates {(0cm,0cm) (0.3cm,0cm)};}, font=\footnotesize}
]
\addplot[vir_6, thick] 
table[x=mu0h,y=b]{data/cell_static_k0p2_jcb_f1_b2.txt};
\addplot[vir_1, thick] 
table[x=mu0h,y=b]{data/cell_static_k1p0_jcb_f1_b2.txt};
\addplot[vir_6, densely dashed, thick] 
table[x=mu0h,y=b]{data/cell_static_k0p2_f1_b2.txt};
\addplot[vir_1, densely dashed, thick] 
table[x=mu0h,y=b]{data/cell_static_k1p0_f1_b2.txt};
\legend{$\mu_0\ku(b) = f_\kappa(b)\, 0.2$ T, $\mu_0\ku(b) = f_\kappa(b)\, 1.0$ T, $\mu_0\ku = 0.2$ T, $\mu_0\ku = 1.0$ T}
\end{axis}
\end{tikzpicture}
\end{subfigure}
\hfill
\begin{subfigure}[b]{0.99\linewidth}  
\centering 
\tikzsetnextfilename{cell_static_f1_b2_m}
\begin{tikzpicture}[trim axis left, trim axis right][font=\small]
\pgfplotsset{set layers}
 \begin{axis}[
tick label style={/pgf/number format/fixed},
width=0.9\textwidth,
height=4.5cm,
grid = major,
grid style = dotted,
ymin=-1.2, 
ymax=1.2,
xmin=-2.2, 
xmax=2.2,
xlabel={$\mu_0 h$ (T)},
ylabel={$\mu_0 m$ (T)},
ylabel style={yshift=-2em},
xlabel style={yshift=0.5em},
xticklabel style={yshift=0.1em},
yticklabel style={xshift=0em},
legend columns=3,
transpose legend,
legend style={at={(0.01, 0.999)}, cells={anchor=west}, anchor=north west, draw=none,fill opacity=0, text opacity = 1, legend image code/.code={\draw[##1,line width=1pt] plot coordinates {(0cm,0cm) (0.3cm,0cm)};}, font=\footnotesize}
]
\addplot[vir_6, densely dashed, thick] 
table[x=mu0h,y=m]{data/cell_static_k0p2_f1_b2.txt};
\addplot[vir_1, densely dashed, thick] 
table[x=mu0h,y=m]{data/cell_static_k1p0_f1_b2.txt};
\addplot[vir_6, thick] 
table[x=mu0h,y=m]{data/cell_static_k0p2_jcb_f1_b2.txt};
\addplot[vir_1, thick] 
table[x=mu0h,y=m]{data/cell_static_k1p0_jcb_f1_b2.txt};
\end{axis}
\end{tikzpicture}
\end{subfigure}
\vspace{-0.2cm}
\caption{Hysteresis curves obtained with one \SC\ with field-dependent or field-independent irreversibility parameters (the function $f_\kappa(b)$ is the same as in Section~\ref{sec_application_static}). The legend is the same for both subfigures. (Up) Magnetic flux density $b$ versus applied field $\mu_0 h$. (Down) Magnetization $\mu_0 m = b-\mu_0h$ versus applied field $\mu_0h$. Note that in this simple case, $h_\text{irr} = - m$.}
    \label{cell_static_f1_b2}
\end{figure}
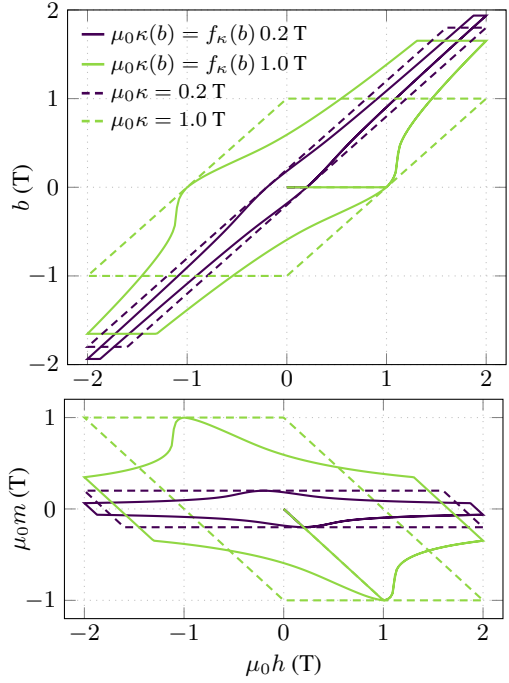

Magnetization loops obtained from \SCs\ with different irreversibility parameters are shown Fig.~\ref{cell_static_f1_b2} for a unidirectional excitation. By construction, these curves are independent of the rate of change of the applied field.

A careful choice of the scaling function $f_\kappa(b)$ allows to reproduce the shape of the lower and upper branches of the major hysteresis loop of Fig.~\ref{magnCurves_2T_uncoupled}. However, with a single \SC, the transitions between these upper and lower branches are not smooth as in the reference solution in Fig.~\ref{magnCurves_2T_uncoupled}, but sharp. Also, for applied fields smaller than $\kappa(0)$, the magnetic flux density remains exactly zero and no loss is generated by the \SC. To obtain smooth transitions and better description at small field amplitudes, combining several cells is necessary. This is discussed in Section~\ref{sec_multiCell}.

\subsection{Composite superconductor cell - \CSC}\label{sec_cell_coupling}

\begin{figure}[h!]
      \begin{subfigure}[b]{0.49\linewidth}  
            \centering 
		\includegraphics[width=0.9\linewidth]{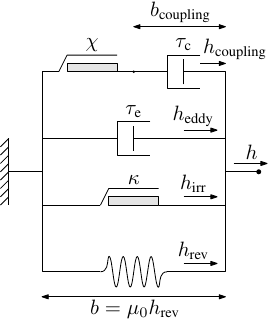}
		\caption{Mechanical analogy.}
		\label{mechanical_analogy_dyn_coupling}	
      \end{subfigure}
      \begin{subfigure}[b]{0.49\linewidth}  
            \centering 
		\includegraphics[width=\textwidth]{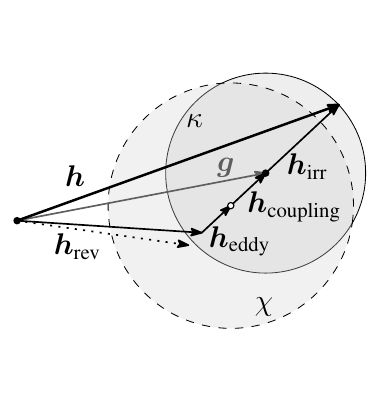}
		\caption{Illustration in 2D.}
		\label{hystModel_dynamic_coupling}	
      \end{subfigure}
\begin{center}
\caption{\CSC\ generalizing the \SC\ by accounting for eddy currents and coupling current effects, as well as for filament coupling effect. Illustration of the magnetic field decomposition into four contributions ($\hrev$, $\hirr$, $\heddy$, and $\hcoupling$). The dashed circle of radius $\kc$ limits the amplitude of the field $\hcoupling$. The dotted arrow represents $\hrev$ at a previous time.}
\label{hystModel_dyn_coupling_combined}
\end{center}
\end{figure}

For regimes in which the superconducting strand response is strongly rate-dependent, the \SC\ presented in the previous section must be generalized. Starting from the \SC, we propose to account for the magnetization due to eddy currents and coupling currents by adding two new contributions in the magnetic field decomposition, $\heddy$ and $\hcoupling$, such that Eq.~\eqref{eq_staticCellHDecomp} is rewritten
\begin{align}\label{eq_dynEddyCouplingCellHDecomp}
\h = \hrev + \hirru + \heddy + \hcoupling.
\end{align}
The equation for $\hirru$ is modified accordingly:
\begin{align}\label{eq_constitutive_dynamic}
\hirru &= \left\{\begin{aligned}
&\h - \vec g,\quad &&\text{if } \|\h - \vec g\| < \ku,\\
&\ku \ \dot \b/\|\dot \b\|,\quad &&\text{if } \|\h - \vec g\| = \ku,\\
\end{aligned}\right.
\end{align}
with $\vec g = \hrev + \heddy + \hcoupling$. The magnetic field decomposition is represented in Fig.~\ref{hystModel_dynamic_coupling}. As described below, the fields $\heddy$ and $\hcoupling$ are always parallel to $\dot \b = \mu_0 \dhrev$. We refer to this new cell as the composite superconductor cell, or \CSC.

Adapting~\cite{henrotte2006dynamical}, the field $\heddy$ is defined by
\begin{align}\label{eq_eddyfield}
\heddy = \teddy \dhrev,
\end{align}
with an eddy current time constant parameter $\teddy$ (s). The eddy component produces a new loss contribution, $\peddy$ (W/m$^3$), which reads
\begin{align}\label{eq_peddy}
\peddy = \heddy \cdot \dot \b = \frac{\mu_0}{\teddy} \|\heddy\|^2 = \frac{\teddy}{\mu_0} \|\dot\b\|^2.
\end{align}

In the mechanical analogy, the inclusion of eddy current effects corresponds to adding a dashpot (viscous friction element) in parallel with the dry friction element and the restoring spring, as illustrated in Fig.~\ref{mechanical_analogy_dyn_coupling}. For slow variations of the force $h$, the dashpot is barely opposing the elongation $b$, whereas for faster variations of $h$, it becomes gradually more blocking and starts to oppose variations of $b$. This reproduces magnetic shielding and increased magnetization due to eddy currents.

The other new contribution $\hcoupling$ is associated with coupling currents~\cite{campbell1982general}. The coupling currents in the conducting matrix are due to net currents flowing in the superconducting filaments, and are therefore also subject to hysteresis and saturation of the filaments. The field $\hcoupling$ has then its own irreversibility parameter $\kc$ (A/m). We define
\begin{align}
\hcoupling &= \left\{\begin{aligned}
&\tcoupling \dhrev, &&\text{if } \|\tcoupling \dhrev\| < \kc,\\
&\kc \ \dhrev/\|\dhrev\|, &&\text{if } \|\tcoupling \dhrev\| \ge \kc,\\
\end{aligned}\right.\label{eq_hirrc}
\end{align}
with a coupling current time constant $\tcoupling$ (s). In order to reproduce reduced magnetization at higher fields, the irreversibility parameter $\chi$ must also be a function of $\b$. We therefore define
\begin{align}\label{eq_jc_scaling_chi}
\kc(\b) = f_\chi(\b) \bar \chi,
\end{align}
with $\bar \chi$ a constant (A/m) and $f_\chi(\b)$ a scaling function. As for $f_\kappa(\b)$, the expression of $f_\chi(\b)$ must be determined in the parameter identification step.

The field $\hcoupling$ introduces two new loss contributions, $\pcoupling$ and  $\pirrc$ (W/m$^3$), that read
\begin{align}
\pcoupling &= \frac{\mu_0}{\tcoupling} \|\hcoupling\|^2 = \frac{\tcoupling}{\mu_0} \|\dbcoupling\|^2,\label{eq_pcoupling}\\
\pirrc &= \hcoupling \cdot \paren{\dot \b -\dbcoupling},\label{eq_pirrc}
\end{align}
with
\begin{align}
\dbcoupling = \frac{\mu_0}{\tcoupling} \hcoupling.
\end{align}

The mechanical analogy associated with the coupling current contribution is represented in Fig.~\ref{mechanical_analogy_dyn_coupling}. The combination of the dashpot in series with the dry friction element reproduces the saturation of the coupling currents. When the force on the dashpot remains below $\kc$, the dry friction element is at rest and produces no hysteresis. This corresponds to a situation in which the coupling currents are small enough for the hysteresis to be neglected. If the force exceeds $\kc$ on the dashpot, it is then subject to hysteresis. The second hysteresis element, with $\kc$, always acts in parallel with the first one, related to the irreversible field $\hirr$. The magnetization of coupled filaments therefore depends on the sum of both, and so is the associated hysteresis loss. We therefore group $\pirr$ and $\pirrc$ into a single component $p_\text{hyst}=\pirr+\pirrc$.

Note that by choosing $\teddy = 0$ and $\tcoupling = 0$, we retrieve the \SC\ of Section~\ref{sec_staticHystCell}. The \SC\ is therefore a particular case of the \CSC.

In summary, the power loss decomposition reads:
\begin{align}\label{eq_ptot_decomposition}
\ptot = \prev + \underbrace{\pirr + \pirrc}_{p_\text{hyst}} + \peddy + \pcoupling,
\end{align}
where $\prev$ corresponds to stored energy, while the other components correspond to dissipated energy.

\begin{figure}[h!]
    \centering
\tikzsetnextfilename{cell_dynamic_fxx_b2_m}
\begin{tikzpicture}[trim axis left, trim axis right][font=\small]
\pgfplotsset{set layers}
 \begin{axis}[
tick label style={/pgf/number format/fixed},
width=\linewidth,
height=7.5cm,
grid = major,
grid style = dotted,
ymin=-2.2, 
ymax=2.2,
xmin=-2.2, 
xmax=2.2,
xlabel={$\mu_0 h$ (T)},
ylabel={$\mu_0 m$ (T)},
ylabel style={yshift=-3em},
xlabel style={yshift=0.5em},
xticklabel style={yshift=0.1em},
yticklabel style={xshift=0em},
yticklabel pos=right,
legend columns=3,
transpose legend,
legend style={at={(0.02, 0.02)}, cells={anchor=west}, anchor=south west, draw=none,fill opacity=0, text opacity = 1, legend image code/.code={\draw[##1,line width=1pt] plot coordinates {(0cm,0cm) (0.3cm,0cm)};}, font=\small}
]
\addplot[vir_6, thick] 
table[x=mu0h,y=m]{data/cell_dynamic_f0p01_b2.txt};
\addplot[vir_5, thick] 
table[x=mu0h,y=m]{data/cell_dynamic_f0p1_b2.txt};
\addplot[vir_3, thick] 
table[x=mu0h,y=m]{data/cell_dynamic_f10_b2.txt};
\addplot[vir_2, thick] 
table[x=mu0h,y=m]{data/cell_dynamic_f100_b2.txt};
\addplot[vir_1, thick] 
table[x=mu0h,y=m]{data/cell_dynamic_f1000_b2.txt};
\addplot[vir_0, thick] 
table[x=mu0h,y=m]{data/cell_dynamic_f10000_b2.txt};
\legend{$f=0.01$ Hz, $f=0.1$ Hz, $f=10$ Hz, $f=10^{2}$ Hz, $f=10^{3}$ Hz, $f=10^{4}$ Hz}
\end{axis}
\end{tikzpicture}
\vspace{-0.2cm}
\caption{Hysteresis curves obtained with one \CSC\ at different frequencies. Cell parameters are: $\mu_0\bar \ku = 0.2$ T, $\mu_0\bar \kc = 0.3$ T, $\tcoupling = 0.2$ s, and $\teddy = 0.3$ ms, and the same functions $f_\kappa(b)$ and $f_\chi(b)$ as in Section~\ref{sec_multiStrand}.}
    \label{cell_coupling_fxx_b1}
\end{figure}
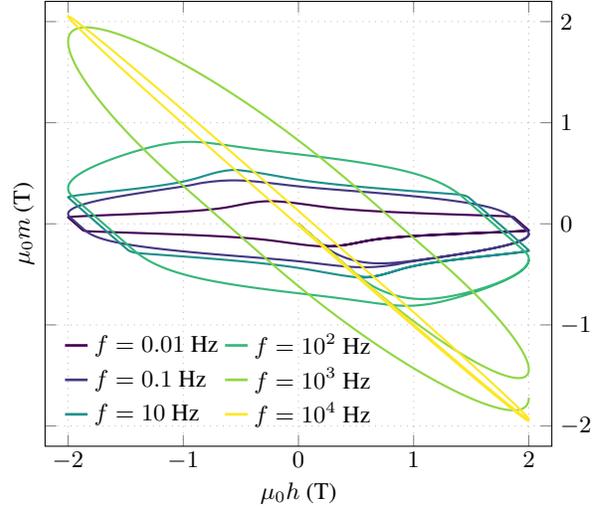

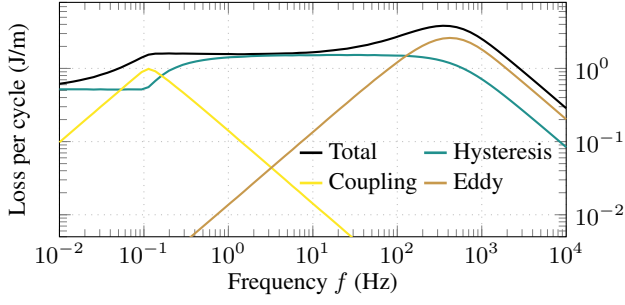
\begin{figure}[h!]
\centering
\tikzsetnextfilename{cell_coupling_lossCycle_f}
\begin{tikzpicture}[trim axis left, trim axis right][font=\small]
\pgfplotsset{set layers}
 	\begin{loglogaxis}[
	tick label style={/pgf/number format/fixed},
    width=\linewidth,
    height=4.7cm,
    grid = major,
    grid style = dotted,
    ymin=5e-3, 
    ymax=8,
    xmin=0.01, 
    xmax=10000,
	xlabel={Frequency $f$ (Hz)},
    ylabel={Loss per cycle (J/m)},
    ylabel style={yshift=-2.2em},
    xlabel style={yshift=0.5em},
    xticklabel style={yshift=0.1em},
    yticklabel style={xshift=0em},
    yticklabel pos=right,
    legend columns=2,
    legend style={at={(1.0, 0.14)}, cells={anchor=west}, anchor=south east, draw=none,fill opacity=0, text opacity = 1, legend image code/.code={\draw[##1,line width=1pt] plot coordinates {(0cm,0cm) (0.3cm,0cm)};}}
    ]
    \addplot[black, thick] 
    table[x=f,y=tot]{data/cell_dynamic_lossCycle_f_per_unit_length.txt};
        \addplot[vir_3, thick] 
    table[x=f,y=fil]{data/cell_dynamic_lossCycle_f_per_unit_length.txt};
        \addplot[vir_0, thick] 
    table[x=f,y=coupling]{data/cell_dynamic_lossCycle_f_per_unit_length.txt};
        \addplot[myorange, thick] 
    table[x=f,y=eddy]{data/cell_dynamic_lossCycle_f_per_unit_length.txt};
    \legend{Total, Hysteresis, Coupling, Eddy}
    \node[anchor=south] at (axis cs: 1000, 6e6) {$\mu_0\hmax = 2$ T};
    \end{loglogaxis}
\end{tikzpicture}%
\vspace{-0.2cm}
\caption{Total loss per cycle and per unit length (loss density multiplied by a surface area $a=\pi\, 0.5^2$ mm$^2$ for comparison with Fig.~\ref{strand_lossCycle}) and individual loss contributions as a function of the frequency $f$, for a magnetic field amplitude of $2$~T obtained with the same \CSC\ as in Fig.~\ref{cell_coupling_fxx_b1}.}
        \label{cell_coupling_lossCycle_f}
\end{figure}

Magnetization curves obtained with one \CSC\ and the associated energy loss per cycle and per unit volume are shown in Figs.~\ref{cell_coupling_fxx_b1} and \ref{cell_coupling_lossCycle_f}. For a given frequency $f$, the energy loss per cycle and per unit volume is computed as
\begin{align}
q_\text{tot} = \int_{1/f}^{2/f} \ptot(t)\, \text{d}t.
\end{align}
The total loss $q_\text{tot}$ is also decomposed in three contributions for easier interpretation, following Eq.~\eqref{eq_ptot_decomposition}.

The hysteresis loss exhibits the expected evolution with frequency: two plateaus at low and middle frequencies, with a transition where the coupling loss peaks, and then a sharp decrease when the eddy loss becomes dominant. Coupling and eddy loss curves follow the shape of two bell curves.

In reality, the eddy current loss is asymptotically decreasing as $1/\sqrt{f}$ at high frequencies due to the skin effect. With the \CSC, one can however only expect the curve to decrease as $1/f$ at high frequencies, as a result of Eq.~\eqref{eq_eddyfield}. For $f\to \infty$, the \CSC\ is therefore not expected to provide perfect results and shall be adapted if one wants better results for very high frequencies ($f \gtrsim 10^3$ Hz).

For the same reasons as for the \SC, several \CSCs\ must be combined to obtain a good description on a range of field amplitudes. This is discussed in the next section.



\subsection{Chain of cells}\label{sec_multiCell}

The \SC\ and the \CSC\ contain the necessary ingredients to reproduce the different regimes observed in a multifilamentary strand, but a single cell alone does not produce a valid response over a wide field range. Indeed, in reality, the evolution towards saturation is smooth and progressive, and not subject to a single threshold field $\ku$ as is the case in Figs.~\ref{cell_static_f1_b2} and \ref{cell_coupling_fxx_b1}.

To better approach the smooth magnetization curves of Figs.~\ref{magnCurves_2T_uncoupled} and \ref{magnCurves_2T}, we can combine several cells into a \textit{chain of cells}. To this end, we follow the approach proposed in~\cite{henrotte2006energy}.

\begin{figure}[h!]
\begin{center}
\includegraphics[width=\linewidth]{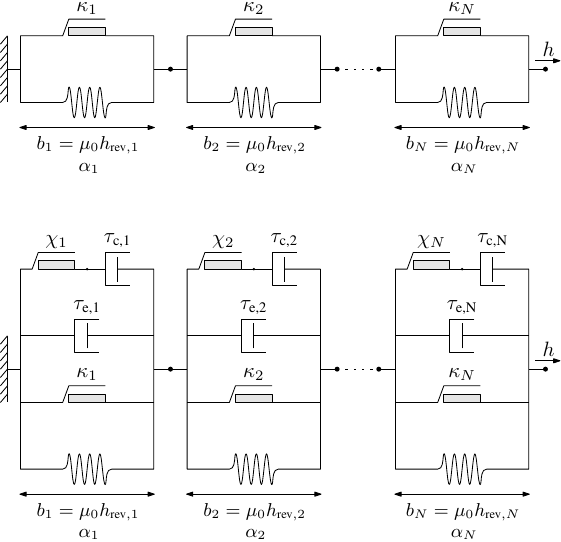}
\caption{(Up) Chain of \SCs. (Down) Chain of \CSCs.}
\label{mechanical_analogy_combined}
\end{center}
\end{figure}

The approach consists in decomposing the total magnetic flux density $\b$ into $N$ contributions $\alpha_k \b_k$, with $k=1,\dots,N$. Each $\b_k$ is described by a distinct hysteresis cell, but all cells are driven by the same magnetic field $\h$. Each cell contributes to the total magnetic flux density with a weight $\alpha_k$. We define
\begin{align}\label{eq_chainEqn}
\b = \sum_{k=1}^N \alpha_k \b_k = \sum_{k=1}^N \alpha_k \mu_0 \hrevk,\ \quad \sum_{k=1}^N \alpha_k = 1,
\end{align}
where $\hrevk$ is the reversible field associated with cell $k$. In the mechanical analogy, this corresponds to connecting cells as a chain, i.e., in series, as illustrated in Fig.~\ref{mechanical_analogy_combined}. 


The power loss is computed by adding contributions of each individual cell, involving the fractions $\alpha_k \dot \b_k$ of the total rate of change $\dot \b$.

The advantage of this approach is that each cell can be solved independently, as they are all subject to the same magnetic field $\h$. Fixed point iterations can be performed to account for field-dependent parameters $\ku$ or $\kc$ that depend on the \textit{total} magnetic flux density $\b$. The equations for each cell are identical to those in Sections~\ref{sec_staticHystCell} and \ref{sec_cell_coupling}, with the only difference that each cell only contributes partly to the magnetic flux density $\b$, with a fraction $\alpha_k \b_k$.

Once parameters are identified (see Section~\ref{sec_paramIdentification}), the chain of cells defines the ROHM model, i.e., the constitutive relationship $\b = \bhyst(\h)$ that reproduces the composite strand magnetization and loss without involving the calculation of any current density distribution.

\section{Implementation and Inclusion in a FE Framework}\label{sec_resolution}

The ROHM model gives the magnetic flux density $\b(t)$ as a function of a given magnetic field $\h(t)$, from which we can compute the magnetization $\m(t) = \b(t)/\mu_0 - \h(t)$ and the power loss. All are vector quantities varying in time. For a numerical simulation, time is discretized and the solutions are computed at successive time steps, based on the knowledge of the solution at previous time steps and on the new value of the magnetic field.

In this section, we show how to solve the two types of cells defined in the previous section, the \SC\ and \CSC, as well as chains of cells. We then describe how to include the ROHM model in a FE framework.

\subsection{\SC}

If the irreversibility parameter $\ku$ is constant, the \SC\ can be solved with an explicit update rule, as proposed in~\cite{henrotte2006energy}. Let $\hrevp$ be the reversible field computed at the previous time step, as a result of the history of $\h$. For a magnetic field $\h$ at the new time step, the reversible field $\hrev$ is updated as follows:
\begin{align}\label{eq_vpm}
\hrev &= \left\{\begin{aligned}
&\hrevp,&&\text{if } \|\h - \hrevp\|\le \ku,\\
&\h - \ku \frac{\h - \hrevp}{\|\h - \hrevp\|},&&\text{otherwise,}
\end{aligned}\right.
\end{align}
The associated magnetic flux density $\b$ is directly given by Eq.~\eqref{eq_bhrev}, and the irreversible field $\hirru$ can be deduced from Eq.~\eqref{eq_staticCellHDecomp}. The instantaneous rates of stored and dissipated energies, $\prev$ and $\pirru$, are readily obtained by Eqs.~\eqref{eq_prev} and \eqref{eq_pirr}, respectively.

If the irreversibility parameter $\ku$ is not constant but a function of $\b$ to account for field-dependent $\jc$, the update rule is no longer explicit. In such a case, an approximate solution can be computed easily with a few fixed point iterations, that is, by solving successively Eq.~\eqref{eq_vpm} with updated values of $\ku(\b) = \ku(\mu_0 \hrev)$ until $\hrev$ does no longer change significantly.

The implementation of the \SC\ is very simple, as it entirely consists in the test of Eq.~\eqref{eq_vpm}. It only requires the knowledge of the previous reversible field $\hrev$, which has to be stored as an internal variable during the numerical simulation.

Note that by contrast with ferromagnetic hysteresis with a nonlinear saturation law~\cite{henrotte2006energy, jacques2018energy}, in this context of superconducting hysteresis, the relationship Eq.~\eqref{eq_bhrev} is linear. The update rule Eq.~\eqref{eq_vpm} is therefore exact~\cite{jacques2018energy}, up to the time discretization error.


\subsection{\CSC}

For the \CSC\ described in Section~\ref{sec_cell_coupling}, the approach is similar. However, two tests are now necessary to account for the two dry friction elements.

We start by updating $\vec g= \hrev+\heddy+\hcoupling$ (see Fig.~\ref{hystModel_dyn_coupling_combined}). Denoting $\vec g_\text{(p)}$ its value at the previous time step, we define the update rule for $\vec g$ as follows:
\begin{align}\label{eq_update_g}
\vec g &= \left\{\begin{aligned}
&\vec g_\text{(p)},\text{ if } \|\h - \vec g_\text{(p)}\|\le \ku \text{ and } \|\h - \hrevp\|\le \ku,\\
&\h - \ku \frac{\h - \hrevp}{\|\h - \hrevp\|},\text{ otherwise,}
\end{aligned}\right.
\end{align}
where the second necessary condition in the first case ensures that $\|\hirr\| < \kappa$ is possible only if $\dot \b = \vec 0$. From $\vec g$, we can compute $\hirru = \h - \vec g$. 


We proceed with the second test, encoded in Eq.~\eqref{eq_hirrc}. We first assume that $\|\tcoupling \dhrev\| \le \chi$ is valid (and we will correct the assumption if it is not the case). If $\|\tcoupling \dhrev\| \le \chi$, then we have the trial fields
\begin{align}
\g &= \hrev^\text{trial} + \heddy^\text{trial} + \hcoupling^\text{trial}\\
&= \hrev^\text{trial} + \teddy \dhrev^\text{trial} + \tcoupling \dhrev^\text{trial}\\
&\approx \hrev^\text{trial} + \frac{\teddy+\tcoupling}{\Delta t} (\hrev^\text{trial} - \hrevp),
\end{align}
with $\hrevp$ the reversible field at the previous time step and $\Delta t$ the current time step size. Solving this equation for $\hrev^\text{trial}$, and then evaluating $\hcoupling^\text{trial} = \tcoupling \dhrev^\text{trial}$ gives the trial value (we drop the $\approx$ sign)
\begin{align}\label{eq_hcoupling_trial}
\hcoupling^\text{trial} = \frac{\tcoupling}{\Delta t + \teddy + \tcoupling} \paren{\vec g - \hrevp}.
\end{align}
If the assumption $\|\hcoupling^\text{trial}\| \le \kc$ is indeed satisfied, then:
\begin{align}
\hcoupling &= \hcoupling^\text{trial},\label{eq_coupling_if_smaller}\\
\heddy &= \frac{\teddy}{ \Delta t + \teddy + \tcoupling} \paren{\vec g - \hrevp}.\label{eq_eddy_if_smaller}
\end{align}
Otherwise, if $\|\hcoupling^\text{trial}\| > \kc$, we have instead (Eq.~\eqref{eq_hirrc}):
\begin{align}
\hcoupling &= \kc\frac{\vec g - \hrevp}{\|\vec g - \hrevp\|},\label{eq_coupling_if_larger}\\
\heddy &= \frac{\teddy}{\Delta t + \teddy} \paren{\vec g - \hcoupling - \hrevp}.\label{eq_eddy_if_larger}
\end{align}
In both cases, the reversible field is obtained via
\begin{align}\label{eq_hrev_for_dyn}
\hrev &= \vec g - \hcoupling - \heddy.
\end{align}
The instantaneous power quantities can be computed by Eqs.~\eqref{eq_prev}, \eqref{eq_pirr}, \eqref{eq_peddy}, \eqref{eq_pcoupling}, and \eqref{eq_pirrc}.

In the case of field-dependent irreversibility parameters $\ku$ and $\kc$, a good convergence is again obtained in a few fixed points iterations, as for the \SC.

The implementation of the \CSC\ is straightforward. It only consists in two tests and a few update equations. The simulation of the \CSC\ requires to keep track of the values of vectors $\vec g$ and $\hrev$ at the previous time step.

\subsection{Chain of cells}

In a ROHM model with a chain of cells as proposed in Section~\ref{sec_multiCell}, each cell $k$ is subject to the same magnetic field $\h$ and produces a distinct magnetic flux density $\b_k$. The total magnetic flux density $\b$ is then computed as a weighted sum of the $\b_k$, as defined in Eq.~\eqref{eq_chainEqn}. This is the only additional step compared to the single cell case.

Apart from this change, each cell can be solved independently exactly as described above, possibly with global fixed point iterations in the case of field-dependent irreversibility parameters $\ku$ and $\kc$ that depend on the \textit{total} magnetic flux density $\b$.

\subsection{Inclusion in a finite element $\phi$-formulation}\label{sec_fem}

The ROHM model $\b = \bhyst(\h)$ can be used as a local constitutive relationship within an FE model.

Let us consider a numerical domain $\O$. It is decomposed into the superconducting strand (or any other superconducting conductor) to be homogenized, denoted as $\Om$, and the complementary domain, denoted as $\Omc$. For simplicity, we assume that $\Omc$ is non-conducting, but the approach can be easily extended to conducting domains as well. Also, in this paper, we only consider cases with no transport current in $\Om$.

The set of equations to solve is therefore:
\begin{equation}\label{magn_equations}
\left\{\begin{aligned}
\div\b &= 0,\\
\curl\h &= \vec 0,
\end{aligned}\right. \quad \text{with} \quad \left\{\begin{aligned} \b &= \bhyst(\h),\quad &\text{in}\ \Om,\\
\b &= \mu_0 \h,\quad &\text{in}\ \Omc.
\end{aligned}\right.
\end{equation}
These equations can be solved numerically with the FE method, on a mesh, i.e., a spatial discretization of the numerical domain $\O$.

As the hysteresis law is driven by the magnetic field $\h$, it is easier to consider an $h$-conform formulation, such as the \hpfOnly~\cite{bossavit1998computational, dular2019finite}. In this particular case with no net currents in $\Om$ and a non-conducting $\Omc$, the \hpf can be reduced to a \pfOnly~\cite{jacques2018energy}. The inclusion of the ROHM model in a $b$-conform formulation is less straightforward and is discussed in~\cite{dular2025finite}.

Starting from an initial solution, the \pf consists in finding a magnetic field $\h = \grad \phi$ with $\phi \in \Phi(\O)$ such that, at subsequent time instants, $\forall \h' = \grad \phi'$ with $\phi' \in\Phi_0(\O)$,
\begin{align}\label{eq_phiformulation}
\volInt{\bhyst(\h)}{\h'}{\Om} + \volInt{\mu_0 \h}{\h'}{\Omc} = 0,
\end{align}
where the notation $\volInt{\vec v}{\vec w}{\O}$ denotes the integral over $\O$ of the dot product of any two vector fields $\vec v$ and $\vec w$. The function space $\Phi(\O)$ contains scalar fields in $H^1(\O)$ that satisfy appropriate essential boundary conditions. The function space $\Phi_0(\O)$ is the same function space but with homogeneous essential boundary conditions. These function spaces are discretized with lowest-order node functions~\cite{bossavit1988whitney}.

The presence of the hysteresis law in Eq.~\eqref{eq_phiformulation} makes the system history-dependent. The fields $\hrev$ and $\vec g$ must therefore be saved at each time step. They are defined in $\Om$ only and are chosen to be element-wise constant vectors, with no continuity constraint across element interfaces.

The hysteresis law also makes the system nonlinear. An iterative scheme is therefore necessary for a numerical simulation. We observed that the Newton-Raphson method leads to very efficient simulations. From an initial iterate $\h_0$, it consists in solving successively a linearized version of Eq.~\eqref{eq_phiformulation} until a given convergence criterion is met. At iteration $i$, the formulation reads, in terms of the unknown field $\h_i$, and the previous (known) iterate $\h_{i-1}$,
\begin{align}
\volInt{\boldsymbol{\mathcal{B}}(\h_{i-1})+\left.\der{\boldsymbol{\mathcal{B}}}{\h}\right|_{\h_{i-1}}\hspace{-0.5cm}\cdot  (\h_i - \h_{i-1})}{\h'}{\Om}&\notag \\
+\volInt{\mu_0 \h_i}{\h'}{\Omc} &= 0,
\end{align}
where the Jacobian tensor $\partial\boldsymbol{\mathcal{B}}/\partial \h$ can be evaluated analytically. Its expression is given in \ref{app_jacobian}. It is not continuous and contains tests as for the constitutive law itself, but it remains very simple to implement. To ensure that the Jacobian tensor is non-singular, it is important that the chain of cells contains at least one cell $k$ with $\ku_k =0$ A/m. Such a cell is usually physically meaningful to represent a volume fraction of the homogenized material that is not subject to hysteresis, such as the normal conducting matrix in the case of a composite strand.

\section{Parameter Identification}\label{sec_paramIdentification}

The ROHM model presented in Section~\ref{sec_cells} provides a flexible tool able to capture a variety of different hysteretic responses thanks to the approach with several cells connected as a chain. The parameter values of each of these cells must be properly chosen in order to reproduce reference solutions. 

In this section, we propose a simple identification procedure for chains of \SCs\ and explain a heuristic for chains of \CSCs, which will be further illustrated in Section~\ref{sec_multiStrand}. The reference solutions are obtained by detailed numerical models such as those presented in Section~\ref{sec_context}.

\subsection{Chain of \SCs}\label{sec_paramIdentification_static}

In a chain of \SCs, there are $2N$ parameters to be defined: the irreversibility parameters $\kappa_k(\b) = f_\kappa(\b) \bar \kappa_k$ and the constant weights $\alpha_k$. The number of cells $N$ must be chosen as a trade-off between computational cost, implementation easiness and accuracy.

\begin{figure}[h!]
\centering
\begin{subfigure}[b]{0.99\linewidth}  
\centering 
\tikzsetnextfilename{static_identification_b}
\begin{tikzpicture}[trim axis left, trim axis right][font=\small]
\pgfplotsset{set layers}
 \begin{axis}[
tick label style={/pgf/number format/fixed},
width=\linewidth,
height=5cm,
grid = major,
grid style = dotted,
ymin=-2.2, 
ymax=2.2,
xmin=-2.1, 
xmax=2.1,
ylabel={$b_\text{in}$ (T)},
ylabel style={yshift=-2.8em},
xlabel style={yshift=0.5em},
xticklabel style={yshift=0.1em},
yticklabel style={xshift=0em},
yticklabel pos=right,
legend columns=3,
transpose legend,
legend style={at={(0.01, 0.0101)}, cells={anchor=west}, anchor=south west, draw=none,fill opacity=0, text opacity = 1}
]
\addplot[vir_6, thick] 
table[x=mu0h,y=b]{data/static_identification_reference_strand_top.txt};
\addplot[vir_2, thick] 
table[x=mu0h,y=b]{data/static_identification_reference_strand_right.txt};
\addplot[vir_2, thick] 
table[x=mu0h,y=b]{data/static_identification_reference_strand_start.txt};
\addplot[vir_2, thick] 
table[x=mu0h,y=b]{data/static_identification_reference_strand_left.txt};
\addplot[vir_6, thick] 
table[x=mu0h,y=b]{data/static_identification_reference_strand_bottom.txt};
\end{axis}
\end{tikzpicture}
\end{subfigure}
\hfill
\vspace{-0.2cm}
\begin{subfigure}[b]{0.99\linewidth}  
\centering 
\tikzsetnextfilename{static_identification}
\begin{tikzpicture}[trim axis left, trim axis right][font=\small]
\pgfplotsset{set layers}
 \begin{axis}[
tick label style={/pgf/number format/fixed},
width=\linewidth,
height=5cm,
grid = major,
grid style = dotted,
ymin=-0.2, 
ymax=0.2,
xmin=-2.1, 
xmax=2.1,
xlabel={$\mu_0h_\text{in}$ (T)},
ylabel={$\mu_0 m$ (T)},
ylabel style={yshift=-2.8em},
xlabel style={yshift=0.5em},
xticklabel style={yshift=0.1em},
yticklabel style={xshift=0em},
yticklabel pos=right,
legend columns=3,
transpose legend,
legend style={at={(0.01, 0.0101)}, cells={anchor=west}, anchor=south west, draw=none,fill opacity=0, text opacity = 1}
]
\addplot[vir_6, thick] 
table[x=mu0h,y=m]{data/static_identification_reference_strand_top.txt};
\addplot[vir_2, thick] 
table[x=mu0h,y=m]{data/static_identification_reference_strand_right.txt};
\addplot[vir_2, thick] 
table[x=mu0h,y=m]{data/static_identification_reference_strand_start.txt};
\addplot[vir_2, thick] 
table[x=mu0h,y=m]{data/static_identification_reference_strand_left.txt};
\addplot[vir_6, thick] 
table[x=mu0h,y=m]{data/static_identification_reference_strand_bottom.txt};
\addplot[gray, densely dashed] 
table[x=mu0h,y=minus_mu0h]{data/static_identification_reference_strand_minusOneLine.txt};
\node[] at (axis cs: -0.4,0.05) {\textcolor{gray}{$b_\text{in} = 0$}};
\draw[<->, gray] (axis cs:1.25, 0) -- (axis cs:1.25, -0.1684);
\draw[-, densely dashed, gray, thin] (axis cs:0.1684, -0.1684) -- (axis cs:1.5, -0.1684);
\node[] at (axis cs: 1.6,-0.13) {\textcolor{gray}{$\mu_0 m_\text{max}$}};
\legend{Step 1, Step 2}
\end{axis}
\end{tikzpicture}
\end{subfigure}
\vspace{-0.2cm}
\caption{Reference magnetization curve for the identification of a chain of \SCs.}
    \label{static_identification}
\end{figure}
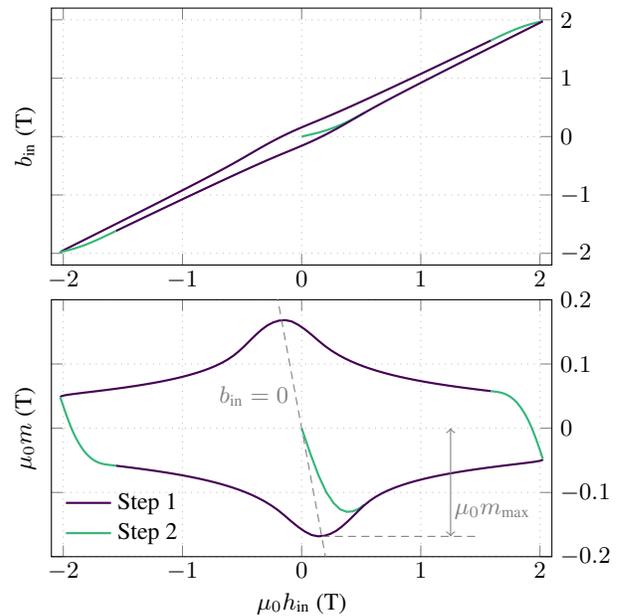

As a reference solution, we consider the magnetization curve of a superconducting strand subject to a unidirectional field $\hs(t) = \hsS(t) \ey = \hmax \sin(2\pi f t) \ey$, with a sufficiently large field $\hmax$ for the system to be fully penetrated, and at a sufficiently low frequency $f$ for the coupling and eddy current effects to be negligible. Based on the reference solution, we compute the average magnetic flux density in the strand $\b_\text{in}$ as 
\begin{align}
\b_\text{in} &= \mu_0\h_\text{in} + \mu_0\m = \mu_0 \paren{\hs + \frac{1}{2} \m},
\end{align}
using Eqs.~\eqref{eq_magnetization} and \eqref{eq_hin_definition}.
An example of the $\b_\text{in}$ and magnetization curves for the composite strand described in Section~\ref{sec_context} is given in Fig.~\ref{static_identification}, with $\mu_0 \hmax = 2$~T and $f=0.01$~Hz. The solution is obtained with a standard 2D FE \hpf without any filament coupling.


The objective is to find the ROHM model parameters such that $\b = \boldsymbol{\mathcal{B}}(\h)$ is as close as possible to $\b_\text{in}$ when $\h = \h_\text{in}$. The proposed identification procedure is decomposed in two steps.

\paragraph{Step 1: Fitting the field-dependent scaling.}

The first step is to fit the irreversibility parameter scaling, $f_\kappa(\b)$, introduced in Eq.~\eqref{eq_jc_scaling} to account for the field-dependent critical current density. For this, we consider parts of the magnetization cycle in which the superconductor is fully magnetized. In Fig.~\ref{static_identification}, this corresponds to the purple curves labelled as Step 1.

For a chain of \SCs, using Eqs.~\eqref{eq_staticCellHDecomp} and~\eqref{eq_chainEqn}, the total magnetization is given by
\begin{align}
\m &=\b/\mu_0 - \h = \sum_{k=1}^N \alpha_k \hrevk - \h\notag\\
&= \sum_{k=1}^N\alpha_k \paren{\hrevk - \h} = \sum_{k=1}^N \alpha_k \hirrk.
\end{align}
In the fully magnetized situation and with a unidirectional excitation, we have
\begin{align}\label{eq_norm_m_SC}
|m| = \sum_{k=1}^N \alpha_k \kappa_k(\b) = f_\kappa(\b) \underbrace{\sum_{k=1}^N \alpha_k \bar\kappa_k}_{m_\text{max}},
\end{align}
with $m_\text{max}$ (A/m) the maximum magnetization. Therefore, the scaling function $f_\kappa(\b)$ directly describes the shape of the purple curves in Fig.~\ref{static_identification}, scaled by $m_\text{max}$.

As a result, the scaling $f_\kappa(\b)$ can be defined as an interpolation function of the purple curves, scaled by $m_\text{max}$. Outside of the field range of the reference solution, $f_\kappa(\b)$ can be extrapolated smoothly, e.g., using the field-dependent critical current density as a scaling. The fact that the model reproduces a maximum magnetization equal to $m_\text{max}$ will be ensured in step 2.


In some circumstances, the maximum magnetization may not be observed at $\b_\text{in} =\vec 0$ exactly. In such a case, the scaling $f_\kappa$ can be written in terms of a weighted sum such as $u\b + (1-u)\mu_0\h$, with $u\in [0,1]$ to be chosen.

Note that choosing $f_\kappa(\b) = \jc(\b)/\jc(\vec 0)$ as a scaling over the whole field range, instead of defining it based on the purple curve as proposed above, does not lead to very accurate results, especially at low fields. The reason is that the strand magnetization depends on the actual field distribution inside it, which is not uniform (but unknown for the homogenized ROHM model), and not only on the average vector $\b_\text{in}$.

\paragraph{Step 2: Choosing the $\bar \kappa_k$ and fitting the weights $\alpha_k$.}

The second step consists in finding the remaining parameters, $\bar \kappa_k$ and $\alpha_k$, in order to reproduce transition branches between fully magnetized states, such as those labelled as Step 2 in Fig.~\ref{static_identification}.

The proposed procedure consists in choosing \textit{a priori} the $N$ values $\bar \kappa_k$ and then fixing the weights $\alpha_k$ accordingly. For simplicity, we arrange the $\bar \kappa_k$ values in increasing order with respect to $k$ and we start with $\bar\kappa_1 = 0$~A/m.

\begin{figure}[h!]
\begin{center}
\includegraphics[width=\linewidth]{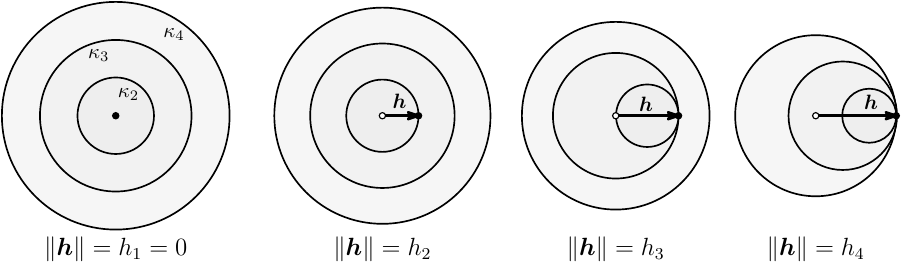}
\caption{Illustration of the magnetic field and a chain of \SCs\ with $N=4$ (the first cell is with $\kappa_1 = 0$ A/m), from virgin state to $\|\h\| = h_4$. Each sphere (circle) represents one cell. The center of the spheres are at positions $\hrevk$, established by the history of $\h$, and their field-dependent radius is equal to $\kappa_k(\b)$. The relative scale is respected compared to Fig.~\ref{static_identification_reference}. Variations of the magnetic field $\h$ that stay inside the smallest non-degenerated sphere produce no loss.}
\label{mechanical_lowField}
\end{center}
\end{figure}

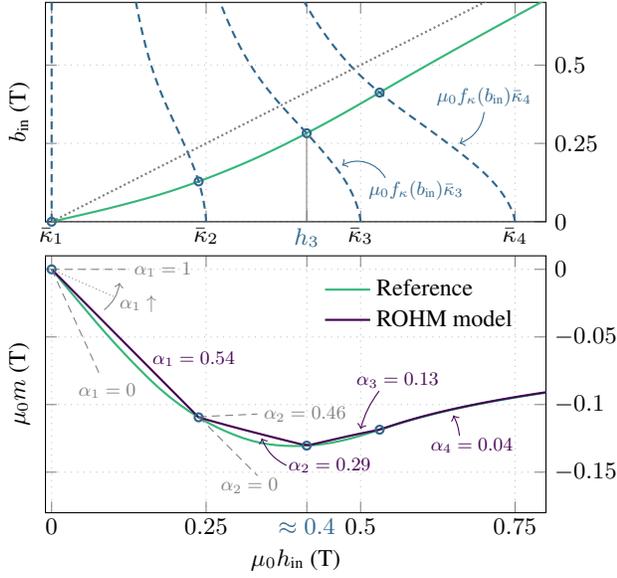
\begin{figure}[h!]
 \begin{subfigure}[b]{0.99\linewidth}  
    \centering
    \tikzsetnextfilename{static_identification_reference_b}
\begin{tikzpicture}[trim axis left, trim axis right][font=\small]
\pgfplotsset{set layers}
 \begin{axis}[
scaled x ticks=false,
xticklabel style={
        /pgf/number format/fixed,
        /pgf/number format/precision=5
},
scaled y ticks=false,
yticklabel style={
        /pgf/number format/fixed,
        /pgf/number format/precision=5
},
width=\linewidth,
height=4.5cm,
grid = major,
grid style = dotted,
ymin=0, 
ymax=0.7,
xmin=-0.01, 
xmax=0.8,
ytick={0, 0.25, 0.5, 0.75},
xtick={0, 0.25, 0.413, 0.5, 0.75},
xticklabels={$\bar\kappa_1$, $\bar\kappa_2$, \textcolor{vir_4}{$h_3$}, $\bar\kappa_3$, $\bar\kappa_4$},
ylabel={$b_\text{in}$ (T)},
ylabel style={yshift=-2.8em},
xlabel style={yshift=0.5em},
xticklabel style={yshift=0.1em},
yticklabel style={xshift=0em},
yticklabel pos=right,
legend style={at={(1, 0.27)}, cells={anchor=west}, anchor=south east, draw=none,fill opacity=0, text opacity = 1}]
\addplot[vir_2, thick] 
table[x=mu0h,y=b]{data/static_identification_reference_strand_new.txt};
\addplot[only marks, color=vir_4, mark=o,mark size=1.5pt, thick] 
table[x=hk,y=bk]{data/static_identification_hkbk_strand_new.txt};
\addplot[vir_4, thick, densely dashed] 
table[x=k2,y=b]{data/static_identification_kappab_strand_new.txt};
\addplot[vir_4, thick, densely dashed] 
table[x=k3,y=b]{data/static_identification_kappab_strand_new.txt};
\addplot[vir_4, thick, densely dashed] 
table[x=k4,y=b]{data/static_identification_kappab_strand_new.txt};
\addplot[vir_4, thick, densely dashed] 
table[x=k1,y=b]{data/static_identification_kappab_strand_new.txt};
\addplot[gray, thick, densely dotted] 
table[x=mu0h,y=mu0h]{data/static_identification_reference_strand.txt};
\node[] at (axis cs: 0.59, 0.1) {\scalebox{0.8}{\textcolor{vir_4}{$\mu_0 f_\kappa(b_\text{in})\bar\kappa_3$}}};
\draw[->, vir_4] (axis cs:0.54, 0.14) to[bend right=45] (axis cs:0.47, 0.18);
\node[] at (axis cs: 0.7, 0.4) {\scalebox{0.8}{\textcolor{vir_4}{$\mu_0 f_\kappa(b_\text{in})\bar\kappa_4$}}};
\draw[->, vir_4] (axis cs:0.7, 0.35) to[bend left=30] (axis cs:0.66, 0.25);
\draw[-, gray, thin] (axis cs:0.4127, 0) -- (axis cs:0.4127, 0.2823);
\end{axis}
\end{tikzpicture}
\end{subfigure}
\hfill\vspace{-0.05cm}
\begin{subfigure}[b]{0.99\linewidth}  
\centering
\tikzsetnextfilename{static_identification_reference_m}
\begin{tikzpicture}[trim axis left, trim axis right][font=\small]
\pgfplotsset{set layers}
 \begin{axis}[
scaled x ticks=false,
xticklabel style={
        /pgf/number format/fixed,
        /pgf/number format/precision=5
},
scaled y ticks=false,
yticklabel style={
        /pgf/number format/fixed,
        /pgf/number format/precision=5
},
width=\linewidth,
height=5cm,
grid = major,
grid style = dotted,
ymin=-0.18, 
ymax=0.01,
xmin=-0.01,
xmax=0.8,
xtick={0, 0.25, 0.413, 0.5, 0.75},
xticklabels={$0$, $0.25$, \textcolor{vir_4}{$\approx 0.4$}, $0.5$, $0.75$},
xlabel={$\mu_0 h_\text{in}$ (T)},
ylabel={$\mu_0 m$ (T)},
ylabel style={yshift=-2.8em},
xlabel style={yshift=0.5em},
xticklabel style={yshift=0.1em},
yticklabel style={xshift=0em},
yticklabel pos=right,
legend columns=3,
transpose legend,
legend style={at={(0.75, 0.95)}, cells={anchor=west}, anchor=north, draw=none,fill opacity=0, text opacity = 1}]
\addplot[vir_2, thick] 
table[x=mu0h,y=m]{data/static_identification_reference_strand_new.txt};
\addplot[vir_6, thick] 
table[x=mu0h,y=m]{data/static_identification_fourCells_strand_new.txt};
\addplot[only marks, color=vir_4, mark=o,mark size=1.5pt, thick] 
table[x=hk,y=mk]{data/static_identification_hkbk_strand_new.txt};
\draw[-, gray, densely dashed, thin] (axis cs:0.0, 0) -- (axis cs:0.125, 0.0);
\draw[-, gray, densely dotted, thin] (axis cs:0.0, 0) -- (axis cs:0.1, -0.02);
\draw[-, gray, densely dashed, thin] (axis cs:0.0, 0) -- (axis cs:0.08, -0.08);
\draw[<-, gray] (axis cs:0.11, -0.01) to[bend left=20] (axis cs:0.085, -0.03);
\node[] at (axis cs: 0.18, 0.0) {\scalebox{0.8}{\textcolor{gray}{$\alpha_1=1$}}};
\node[] at (axis cs: 0.14, -0.025) {\scalebox{0.8}{\textcolor{gray}{$\alpha_1 \uparrow$}}};
\node[] at (axis cs: 0.09, -0.09) {\scalebox{0.8}{\textcolor{gray}{$\alpha_1=0$}}};
\node[] at (axis cs: 0.23, -0.066) {\scalebox{0.8}{\textcolor{vir_6}{$\alpha_1=0.54$}}};

\draw[-, gray, densely dashed, thin] (axis cs:0.238, -0.109) -- (axis cs:0.3332, -0.1526);
\draw[-, gray, densely dashed, thin] (axis cs:0.238, -0.109) -- (axis cs:0.333, -0.107);
\node[] at (axis cs: 0.41, -0.107) {\scalebox{0.8}{\textcolor{gray}{$\alpha_2= 0.46$}}};
\node[] at (axis cs: 0.32, -0.16) {\scalebox{0.8}{\textcolor{gray}{$\alpha_2=0$}}};
\node[] at (axis cs: 0.45, -0.145) {\scalebox{0.8}{\textcolor{vir_6}{$\alpha_2=0.29$}}};
\draw[->, vir_6] (axis cs:0.38, -0.145) to[bend left=20] (axis cs:0.34, -0.125);
\node[] at (axis cs: 0.56, -0.082) {\scalebox{0.8}{\textcolor{vir_6}{$\alpha_3=0.13$}}};
\draw[->, vir_6] (axis cs:0.53, -0.09) to[bend right=10] (axis cs:0.5, -0.118);
\node[] at (axis cs: 0.68, -0.13) {\scalebox{0.8}{\textcolor{vir_6}{$\alpha_4=0.04$}}};
\draw[->, vir_6] (axis cs:0.66, -0.123) to[bend right=10] (axis cs:0.65, -0.108);
\legend{Reference, ROHM model}
\end{axis}
\end{tikzpicture}
\end{subfigure}
\vspace{-0.2cm}
\caption{Parameter identification procedure for a chain of S cells. Example with $N=4$ cells, with $\mu_0\kappa_k = (k-1)\, 0.25$~T. (Up) The dotted gray line represents $b_\text{in}=\mu_0 h_\text{in}$. The dashed curves represent the field-dependent irreversibility parameters. Their intersection with the reference curve defines the threshold values $h_k$. (Down) Illustration of the influence of the weights $\alpha_k$ on the ROHM model solution. The dashed gray curves depict what the solution would partially look like for different values of $\alpha_k$.}
    \label{static_identification_reference}
\end{figure}

We consider the virgin magnetization curve represented by the green curve starting at origin in Fig.~\ref{static_identification}. Initially, $h_{\text{rev},k} = 0$, $\forall k= 1, \dots, N$. As the magnetic field $h_\text{in}$ increases, it reaches the surface of the cell spheres one by one (see Figs.~\ref{hystModel_static} and \ref{mechanical_lowField}). This happens at successive threshold field values that depend on the $\bar \kappa_k$ constants and on the $f_\kappa(b_\text{in})$ scaling determined in Step 1. We denote these threshold fields as $h_k$, as illustrated in Fig.~\ref{static_identification_reference}.

For $h<h_{k}$, we have $h_{\text{rev},j} = 0$, $\forall j = k, \dots, N$, such that only the first $k-1$ cells contribute to the magnetic flux density $b$. The weights $\alpha_{k-1}$ can therefore be consecutively adjusted by forcing the solution of the ROHM model to match the reference solution at each $h_k$, for $k=2,\dots,N$. This adjustment can be automated with a simple iterative routine. The influence of the weights $\alpha_k$ on the ROHM model solution is illustrated in Fig.~\ref{static_identification_reference}. The last weight $\alpha_N$ is calculated so that all weights add up to one.


The choice of the number of cells $N$ and the \textit{a priori} distribution of values for the $\bar \kappa_k$ depend on the desired accuracy and application. This is discussed in Section~\ref{sec_application_static}.



\subsection{Chain of \CSCs}

For a chain of \CSCs, there are $5N$ parameters to be defined: $\ku_k$, $\alpha_k$, $\tek$, $\tck$, and $\kc_k$, for $k=1,\dots,N$. To identify all these parameters, reference solutions at different frequencies are necessary. We decompose the identification procedure in three steps.

\paragraph{Step 1: Reproducing the rate-independent regime.}
A reference solution at a sufficiently low frequency, so that the eddy and coupling current effects are negligible, can be used exactly as described in the previous section to define the $\ku_k$ (with both the $f_\kappa(\b)$ scaling and the $\bar \ku_k$ values) and $\alpha_k$. This leads to a number of $3N$ remaining parameters.

\paragraph{Step 2: Fitting the time constants.}
The time constants $\tek$ and $\tck$ can be chosen so that the positions of the peaks in eddy and coupling losses correspond to those of reference solutions, such as shown in Fig.~\ref{strand_lossCycle}. Choosing identical values for all the cells already leads to good results within limited amplitude ranges. To reproduce the observation that the maxima in coupling and eddy loss are observed at lower frequencies when the field amplitude increases, one can choose larger values of $\tek$ and $\tck$ for cells associated with larger irreversibility parameters. This will be illustrated in Section~\ref{sec_multiStrand}.

\paragraph{Step 3: Reproducing the coupled filament magnetization.}
Once the time constants are fixed, the only remaining parameters are the $\kc_k(\b) = f_\chi(\b) \bar \kc_k$. The scaling $f_\chi(\b)$ can be defined as in the static case (Step 1), based on a reference solution at a sufficiently high frequency for the dynamic effects to be visible in the magnetization curve, e.g., at $f= 100$~Hz for the 54-filament strand, as shown in Fig.~\ref{magnCurves_2T_hin}. Then, the values $\bar \kc_k$ can be identified in order to best reproduce the transition branches, as in Step 2 of the static case, but now with values $\bar \kc_k$ as unknowns instead of the weights $\alpha_k$ which are already fixed.

\section{Results with a chain of \SCs}\label{sec_application_static}

The first application consists of the 54-filament strand presented in Section~\ref{sec_context}, subject to a transverse field varying sufficiently slowly for the eddy and coupling current effects to be negligible. We choose a frequency $f= 0.01$~Hz. For the reference solution, we assume a non-conducting matrix and use a classical \hpfOnly~\cite{bossavit1998computational}, so that coupling effects are not present.

Because of the finite value of the power index $n=30$, the strand response is not truly rate-independent (this would only be the case at the limit $n\to \infty$). Still, we model the macroscopic strand response (magnetization and loss) with a chain of \SCs\ as a first approximation, and show that it already provides very good results.


We start the analysis by identifying the parameters of the chain of \SCs. We then compare the prediction of the resulting model for different types of excitations, both unidirectional and bidirectional.

\subsection{Parameter identification}\label{sec_application_static_identification}

\begin{figure}[!h]
\begin{subfigure}[b]{0.45\linewidth}
\small
\centering
\begin{tabular}{l r r}
\hline
$k$ & \multicolumn{1}{c}{$\mu_0\bar\kappa_k$} & \multicolumn{1}{c}{$\alpha_k$} \\
& \multicolumn{1}{c}{(mT)} & \multicolumn{1}{c}{($\%$)}\\
\hline
$1$ & $0$ & $48.24$ \\
\hline
$2$ & $150$ & $18.07$  \\
$3$ & $300$ & $16.99$  \\
$4$ & $450$ & $9.31$   \\
$5$ & $600$ & $5.16$   \\
$6$ & $750$ & $2.23$   \\
\hline
\end{tabular}
\end{subfigure}
\begin{subfigure}[b]{0.54\linewidth}
\centering
\tikzsetnextfilename{f_kappa}
\begin{tikzpicture}[][font=\small]
\pgfplotsset{set layers}
 \begin{axis}[
tick label style={/pgf/number format/fixed},
width=\textwidth,
height=4.5cm,
grid = major,
grid style = dotted,
ymin=0, 
ymax=1.05,
xmin=0, 
xmax=3.1,
ytick={0, 0.2, 0.4, 0.6, 0.8, 1.0},
xlabel={$b$ (T)},
ylabel={Scaling function (-)},
ylabel style={yshift=-1.5em},
xlabel style={yshift=0.6em},
xticklabel style={yshift=0.1em},
yticklabel style={xshift=0em},
legend columns=4,
transpose legend,
legend style={at={(0.99, 0.96)}, cells={anchor=west}, anchor=north east, draw=none,fill opacity=0, text opacity = 1}
]
\addplot[vir_6, thick] 
table[x=b,y=fk]{data/f_kappa.txt};
\legend{$f_\kappa(b)$}
\end{axis}
\end{tikzpicture}
\vspace{-2cm}
\end{subfigure}
\vspace{0.2cm}
\caption{(Left) Parameters of a chain of \SCs\ with $N = 6$, identified on the reference magnetization loop shown in Fig.~\ref{static_identification}. (Right) Scaling function $f_\kappa(b)$.}
\label{tab_param_static}
\end{figure}
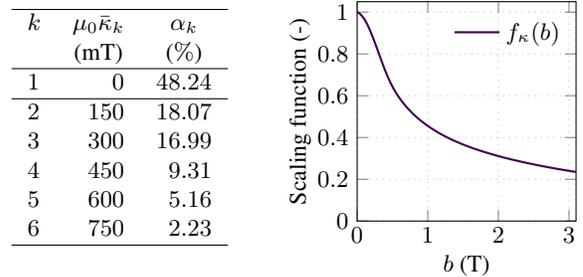

The reference solution is the major loop represented in Fig.~\ref{static_identification}. We choose $N=6$ and apply the two-step procedure described in Section~\ref{sec_paramIdentification_static} with $N$ equally spaced values $\mu_0\bar \kappa_k$ from $0$~mT to $750$~mT. The obtained parameters are given in Fig.~\ref{tab_param_static}.


It is interesting to notice that the first \SC, which is anhysteretic ($\kappa_1 = 0$ A/m), almost contributes to half of the magnetic flux density. This is due to the large fraction of normal conductor in the strand, the copper matrix (copper fraction in the cross section is $56\%$), which does not behave as a hysteretic material.

\begin{figure*}[h]
\begin{subfigure}[b]{0.49\linewidth}
\centering
\tikzsetnextfilename{static_result_majorLoop_magn}
\begin{tikzpicture}[trim axis left, trim axis right][font=\small]
\begin{groupplot}[group style={group size=1 by 3,
       horizontal sep=0pt,
       vertical sep=1cm},
     ] 
\tikzsetnextfilename{static_result_majorLoop_source}
\nextgroupplot[
tick label style={/pgf/number format/fixed},
width=\textwidth,
height=3.5cm,
grid = major,
grid style = dotted,
ymin=-2.6, 
ymax=2.6,
xmin=0, 
xmax=125,
xlabel={Time (s)},
ylabel={$\mu_0 h_\text{in}$ (T)},
ylabel style={yshift=-3em},
xlabel style={yshift=0.5em},
xticklabel style={yshift=0.1em},
yticklabel style={xshift=0em},
yticklabel pos=right,
legend style={at={(0.01, 0.99)}, cells={anchor=west}, anchor=north west, draw=none,fill opacity=0, text opacity = 1}
]
\addplot[myorange, thick] 
table[x=t,y=mu0h]{data/static_result_majorLoop_strand_N6.txt};
\nextgroupplot[
scaled y ticks=false,
yticklabel style={
        /pgf/number format/fixed,
        /pgf/number format/precision=5
},
tick label style={/pgf/number format/fixed},
width=\textwidth,
height=5.5cm,
grid = major,
grid style = dotted,
xmin=-2.6, 
xmax=2.6,
ymin=-0.2, 
ymax=0.2,
xlabel={$\mu_0 h_\text{in}$ (T)},
ylabel={$\mu_0 m$ (T)},
ylabel style={yshift=-3em},
xlabel style={yshift=0.5em},
xticklabel style={yshift=0.1em},
yticklabel style={xshift=0em},
yticklabel pos=right,
legend columns=3,
transpose legend,
colormap name=vir_map,
colorbar/width=2.5mm,
legend style={at={(0.0101, 0.990)}, cells={anchor=west}, anchor=north west, draw=none,fill opacity=0, text opacity = 1}
]
\addplot[vir_2, densely dashed, thick] 
table[x=mu0h,y=mref]{data/static_result_majorLoop_strand_N6.txt};
\addplot[mesh, line legend, thick, point meta max = 20, point meta=explicit] 
table[x=mu0h, y=mmodel, meta=pmodel]{data/static_result_majorLoop_strand_N6.txt};
\legend{Reference, ROHM}
\tikzsetnextfilename{static_result_majorLoop_power}
\nextgroupplot[
tick label style={/pgf/number format/fixed},
width=\textwidth,
height=4.5cm,
grid = major,
grid style = dotted,
ymin=0, 
xmin=0, 
xmax=125,
xlabel={Time (s)},
ylabel={Power loss (mW/m)},
ylabel style={yshift=-3em},
xlabel style={yshift=0.5em},
xticklabel style={yshift=0.1em},
yticklabel style={xshift=0em},
yticklabel pos=right,
legend style={at={(0.010, 0.99)}, cells={anchor=west}, anchor=north west, draw=none,fill opacity=0, text opacity = 1}
]
\addplot[vir_2, densely dashed, thick] 
table[x=t,y=pref]{data/static_result_majorLoop_strand_N6.txt};
\addplot[mesh, line legend, thick, point meta max = 20, point meta=explicit]
table[x=t,y=pmodel, meta=pmodel]{data/static_result_majorLoop_strand_N6.txt};
\legend{Reference, ROHM}
\end{groupplot}
\end{tikzpicture}
\caption{Major loop $\hsS^{(1)}(t)$.}
\end{subfigure}
    \centering
\begin{subfigure}[b]{0.49\linewidth}    
\centering
\tikzsetnextfilename{static_result_minorLoops_magn}
\begin{tikzpicture}[trim axis left, trim axis right][font=\small]
\pgfplotsset{set layers}
\begin{groupplot}[group style={group size=1 by 3,
       horizontal sep=0pt,
       vertical sep=1cm},
     ] 
     \tikzsetnextfilename{static_result_minorLoop_source}
\nextgroupplot[
tick label style={/pgf/number format/fixed},
width=\textwidth,
height=3.5cm,
grid = major,
grid style = dotted,
ymin=-2.6, 
ymax=2.6,
xmin=0, 
xmax=250,
xlabel={Time (s)},
ylabel={$\mu_0 h_\text{in}$ (T)},
ylabel style={yshift=-3em},
xlabel style={yshift=0.5em},
xticklabel style={yshift=0.1em},
yticklabel style={xshift=0em},
yticklabel pos=right,
legend style={at={(0.01, 0.99)}, cells={anchor=west}, anchor=north west, draw=none,fill opacity=0, text opacity = 1}
]
\addplot[myorange, thick] 
table[x=t,y=mu0h]{data/static_result_minorLoop_strand_N6.txt};
\nextgroupplot[
scaled y ticks=false,
yticklabel style={
        /pgf/number format/fixed,
        /pgf/number format/precision=5
},
tick label style={/pgf/number format/fixed},
width=\textwidth,
height=5.5cm,
grid = major,
grid style = dotted,
ymin=-0.2, 
ymax=0.2,
xmin=-2.6, 
xmax=2.6,
xlabel={$\mu_0 h_\text{in}$ (T)},
ylabel={$\mu_0 m$ (T)},
ylabel style={yshift=-3em},
xlabel style={yshift=0.5em},
xticklabel style={yshift=0.1em},
yticklabel style={xshift=0em},
yticklabel pos=right,
legend columns=3,
transpose legend,
colormap name=vir_map,
legend style={at={(0.010, 0.990)}, cells={anchor=west}, anchor=north west, draw=none,fill opacity=0, text opacity = 1}
]
\addplot[vir_2, densely dashed, thick] 
table[x=mu0h,y=mref]{data/static_result_minorLoop_strand_N6.txt};
\addplot[mesh, line legend, thick, point meta max = 20, point meta=explicit] 
table[x=mu0h, y=mmodel, meta=pmodel]{data/static_result_minorLoop_strand_N6.txt};
\legend{Reference, ROHM}
\tikzsetnextfilename{static_result_minorLoops_power}
\nextgroupplot[
tick label style={/pgf/number format/fixed},
width=\textwidth,
height=4.5cm,
grid = major,
grid style = dotted,
ymin=0, 
xmin=0, 
xmax=250,
xlabel={Time (s)},
ylabel={Power loss (mW/m)},
ylabel style={yshift=-3em},
xlabel style={yshift=0.5em},
xticklabel style={yshift=0.1em},
yticklabel style={xshift=0em},
yticklabel pos=right,
legend style={at={(0.0101, 0.99)}, cells={anchor=west}, anchor=north west, draw=none,fill opacity=0, text opacity = 1}
]
\addplot[vir_2, densely dashed, thick] 
table[x=t,y=pref]{data/static_result_minorLoop_strand.txt};
\addplot[mesh, line legend, thick, point meta max = 20, point meta=explicit]
table[x=t,y=pmodel, meta=pmodel]{data/static_result_minorLoop_strand.txt};
\coordinate (insetPosition) at (rel axis cs:0.2,0.4);
\legend{Reference, ROHM}
\end{groupplot}
\end{tikzpicture}
\caption{Minor loops $\hsS^{(2)}(t)$.}
\end{subfigure}
\caption{Comparison of the ROHM model results with reference solutions. Unidirectional transverse magnetic field. Chain of \SCs\ with $N=6$ cells, parameters of Fig.~\ref{tab_param_static}. (a) Harmonic excitation $\hsS^{(1)}(t)$. (b) Biharmonic excitation $\hsS^{(2)}(t)$. (Up) Internal magnetic field, the input of the ROHM model. (Middle) Magnetization loop. The curves of the ROHM model are colored as a function of the instantaneous power, according to the values in the bottom plots. (Down) Dissipated power per unit length as a function of time, equal to $a\,p_\text{irr}(t)$, with $a = \pi d^2/4$ the surface area of the strand cross section.}
    \label{static_result_unidirectional}
\end{figure*}
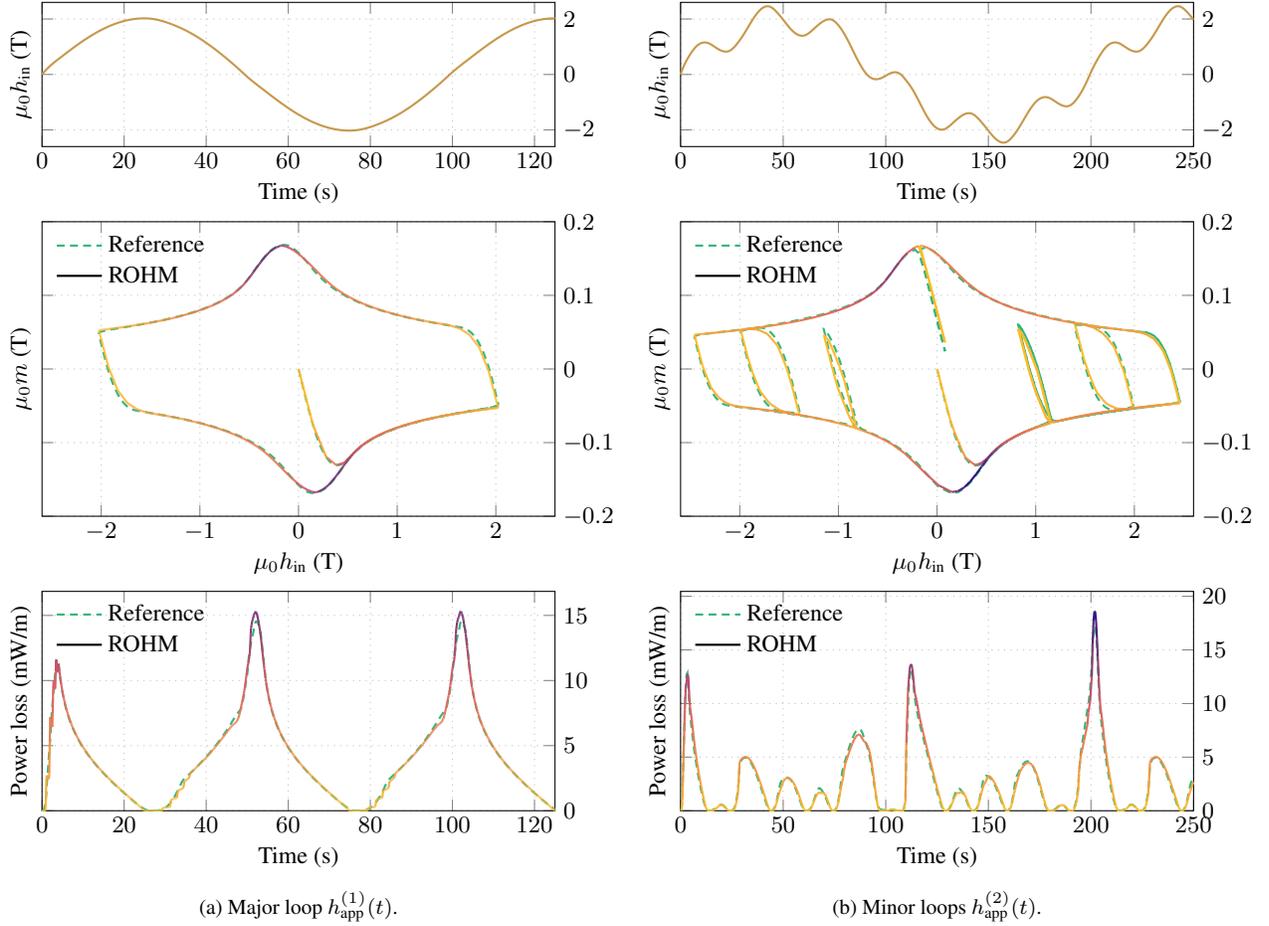

\subsection{Unidirectional excitation}

Using the parameters obtained above, we compare the predictions of the ROHM model in terms of magnetization and loss for two types of unidirectional excitations along $\ey$. The first one is a harmonic field:
\begin{align}
\mu_0 \hsS^{(1)}(t) &= \mu_0 \hmax \sin(2\pi ft),
\end{align}
with $\mu_0 \hmax =  2$~T and $f = 0.01$~Hz. The second one is a biharmonic field, resulting in a response with minor magnetization loops:
\begin{align}
\mu_0 \hsS^{(2)}(t) &= \mu_0 \hmax \paren{\sin(\pi f t) + 0.25 \sin(6\pi f t)}.
\end{align}

Results are given in Fig.~\ref{static_result_unidirectional}. With $N=6$ cells, the ROHM model reproduces very well both major and minor magnetization loops. The instantaneous loss is remarkably well reproduced in both cases, with a relative error on the total loss below $1\%$.

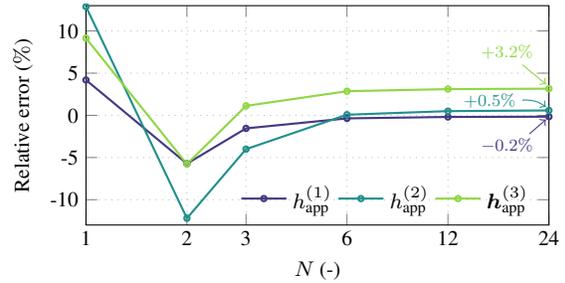
\begin{figure}[h!]
    \centering
\centering
\tikzsetnextfilename{fig_static_N_effect}
\begin{tikzpicture}[trim axis left, trim axis right][font=\footnotesize]
\pgfplotsset{set layers}
 \begin{semilogxaxis}[
scaled ticks=false,
tick label style={/pgf/number format/fixed},
width=0.93\linewidth,
height=4.5cm,
grid = major,
grid style = dotted,
ymin=-13, 
ymax=13,
xmin=1, 
xmax=24,
xtick={1, 2, 3, 6, 12, 24},
xticklabels={1, 2, 3, 6, 12, 24},
ytick={-10, -5, 0, 5, 10},
yticklabels={-10, -5, 0, 5, 10},
xlabel={$N$ (-)},
ylabel={Relative error ($\%$)},
ylabel style={yshift=-1.4em},
xlabel style={yshift=0.5em},
xticklabel style={yshift=0.1em},
yticklabel style={xshift=0em},
legend columns=3,
legend style={at={(0.65, 0.01)}, cells={anchor=west}, anchor=south, draw=none,fill opacity=0, text opacity = 1}
]
\addplot[vir_5, thick, mark=o, mark size=1pt] 
    coordinates {(1,4.19) (2, -5.73) (3, -1.546) (6, -0.36) (12, -0.19) (24, -0.16)};
\addplot[vir_3, thick, mark=o, mark size=1pt] 
    coordinates {(1,12.88) (2, -12.2) (3, -4.01) (6, 0.075) (12, 0.5) (24, 0.579)};
\addplot[vir_1, thick, mark=o, mark size=1pt] 
    coordinates {(1,9.15) (2, -5.77) (3, 1.12) (6, 2.86) (12, 3.11) (24, 3.16)};
\node[] at (axis cs: 18, -3.5) {\scalebox{0.8}{\textcolor{vir_5}{$-0.2\%$}}};
\draw[->, vir_5] (axis cs:21, -2.2) to (axis cs:23.5, -0.5);
\node[] at (axis cs: 16, 1.8) {\scalebox{0.8}{\textcolor{vir_3}{$+0.5\%$}}};
\draw[->, vir_3] (axis cs:20, 1.8) to[bend left=20] (axis cs:23.5, 1);
\node[] at (axis cs: 18, 7.5) {\scalebox{0.8}{\textcolor{vir_1}{$+3.2\%$}}};
\draw[->, vir_1] (axis cs:20, 6) to (axis cs:23.5, 3.7);
\legend{$\hsS^{(1)}$, $\hsS^{(2)}$, $\hs^{(3)}$}
\end{semilogxaxis}
\end{tikzpicture}
\vspace{-0.2cm}
\caption{Evolution of the relative error $(Q_\text{hyst} - Q_\text{ref})/Q_\text{ref}$ between total losses predicted by the ROHM model ($Q_\text{hyst}$) and the reference model ($Q_\text{ref}$), for different excitations, as a function of the number of cells $N$, with equally spaced values $\mu_0 \bar \kappa_k$, from $0$~T to $0.75$~T. In the case $N=1$, $\mu_0\bar \kappa_1 = 0.168$ T and $\alpha_1 = 1$.}
    \label{fig_static_N_effect}
\end{figure}

The influence of the number of cells on the relative error on the total loss is shown in Fig.~\ref{fig_static_N_effect}. Increasing the number of cells helps reducing the error for low $N$ values ($\le 3$). The error then stabilizes (for $N\ge 6$) to a small, non-zero value. 


The major advantage of the method is its computational efficiency. Once the model parameters are identified, solving the ROHM model is tremendously faster than solving a detailed FE model. Typically a few milliseconds only is required for the ROHM model compared to a few tens of minutes for the 54-filament reference FE simulation. This efficiency is crucial in view of simulating full-scale superconducting systems.

In practice, the number of cells and the distribution of $\bar \kappa_k$ values must be chosen depending on the actual excitations to be considered in practice. As already mentioned, a chain of \SCs\ produces no loss for field variations smaller than the smallest non-zero irreversibility parameter $\kappa_k(\b)$, as illustrated in Fig.~\ref{mechanical_lowField}. To reproduce accurately power loss at low fields, or for low field ripples, the chain of \SCs\ must therefore contain cells with sufficiently small irreversibility parameters. The same comment holds for a chain of \CSCs, and will be further illustrated in Section~\ref{sec_multiStrand}.



\subsection{Bidirectional excitation}\label{sec_rotating_static}

\begin{figure}[h]
\begin{subfigure}[t]{0.49\linewidth}
\centering
\tikzsetnextfilename{static_result_rotational_app}
\begin{tikzpicture}[trim axis left, trim axis right][font=\small]
\begin{axis}[
scaled y ticks=false,
yticklabel style={
        /pgf/number format/fixed,
        /pgf/number format/precision=5
},
tick label style={/pgf/number format/fixed},
width=1.1\textwidth,
height=5cm,
grid = major,
grid style = dotted,
xmin=-0.2, 
xmax=2.2,
axis equal,
ylabel style={yshift=-1em},
xlabel style={yshift=0.5em},
xticklabel style={yshift=0.1em},
yticklabel style={xshift=0em},
legend columns=3,
transpose legend,
legend style={at={(0.01, 0.01)}, cells={anchor=west}, anchor=south west, draw=none,fill opacity=0, text opacity = 1},]
\addplot[myorange, thick, mark=none] 
table[x=mu0hx,y=mu0hy]{data/static_result_strand_rotational_N6.txt};
\addplot[scatter, only marks, scatter/classes={0={bw_6}, 1={bw_5}, 2={bw_4}, 3={bw_3}, 4={bw_2}, 5={bw_1}},
        scatter src=explicit symbolic, mark=*, thick, mark size=0.8pt] 
table[x=mu0hx,y=mu0hy, meta=style]{data/static_result_strand_rotational_N6_few.txt};
\draw[->] (axis cs:0, 0) -- (axis cs:1, 0);
\draw[->] (axis cs:0, 0) -- (axis cs:0, 1);
\node[] at (axis cs: 1,-0.2) {\footnotesize $\mu_0 h_{\text{in},x}$};
\node[] at (axis cs: 0.2,1.14) {\footnotesize $\mu_0 h_{\text{in},y}$};
\legend{Inner field}
\end{axis}
\end{tikzpicture}
\end{subfigure}
    \hspace{-0.9cm}
\begin{subfigure}[t]{0.49\linewidth}    
\centering
\tikzsetnextfilename{static_result_rotational_magn_2}
\begin{tikzpicture}[trim axis left, trim axis right][font=\small]
\pgfplotsset{set layers}
\begin{axis}[
scaled y ticks=false,
yticklabel style={
        /pgf/number format/fixed,
        /pgf/number format/precision=5
},
scaled x ticks=false,
xticklabel style={
        /pgf/number format/fixed,
        /pgf/number format/precision=5
},
tick label style={/pgf/number format/fixed},
width=1.15\textwidth,
height=5cm,
grid = major,
grid style = dotted,
axis equal,
xmin=-0.1, 
xmax=0.12,
ymin=-0.2, 
ymax=0.12,
ylabel style={yshift=-3em},
xlabel style={yshift=0.5em},
xticklabel style={yshift=0.1em},
yticklabel style={xshift=0em},
yticklabel pos=right,
legend columns=3,
transpose legend,
legend style={at={(0.01, 1)}, cells={anchor=west}, anchor=north west, draw=none,fill opacity=0, text opacity = 1}
]
\addplot[vir_2, densely dashed, thick] 
table[x=mxref,y=myref]{data/static_result_strand_rotational_N6.txt};
\addplot[mesh, line legend, thick, point meta max = 7.5, point meta=explicit]
table[x=mxhyst,y=myhyst, meta=pmodel]{data/static_result_strand_rotational_N6.txt};
\addplot[scatter, only marks, mark=*, scatter/classes={0={bw_6}, 1={bw_5}, 2={bw_4}, 3={bw_3}, 4={bw_2}, 5={bw_1}}, scatter src=explicit symbolic, thick, mark size=0.8pt] 
table[x=mxref,y=myref, meta=style]{data/static_result_strand_rotational_N6_few.txt};
\addplot[scatter, only marks, mark=x, scatter/classes={0={bw_6}, 1={bw_5}, 2={bw_4}, 3={bw_3}, 4={bw_2}, 5={bw_1}}, scatter src=explicit symbolic, mark size=2pt] 
table[x=mxhyst,y=myhyst, meta=style]{data/static_result_strand_rotational_N6_few.txt};
\draw[->] (axis cs:0, 0) -- (axis cs:0.1, 0);
\draw[->] (axis cs:0, 0) -- (axis cs:0, 0.1);
\node[] at (axis cs: 0.11,0.013) {\footnotesize $\mu_0 m_{x}$};
\node[] at (axis cs: 0.04,0.085) {\footnotesize $\mu_0 m_{y}$};
\tikzsetnextfilename{static_result_rotational_power}
\end{axis}
\end{tikzpicture}
\end{subfigure}
 \centering
\begin{subfigure}[b]{\linewidth}    
\centering
\tikzsetnextfilename{static_result_rotational_power}
\begin{tikzpicture}[trim axis left, trim axis right][font=\small]
\begin{axis}[
tick label style={/pgf/number format/fixed},
width=\textwidth,
height=4.5cm,
grid = major,
grid style = dotted,
ymin=0, 
xmin=0, 
xmax=125,
xlabel={Time (s)},
ylabel={Power loss (mW/m)},
ylabel style={yshift=-3em},
xlabel style={yshift=0.5em},
xticklabel style={yshift=0.1em},
yticklabel style={xshift=0em},
yticklabel pos=right,
legend columns=2,
transpose legend,
legend style={at={(0.4, 0.95)}, cells={anchor=west}, anchor=north, draw=none,fill opacity=0, text opacity = 1}
]
\addplot[vir_2, densely dashed, thick] 
table[x=t,y=pref]{data/static_result_strand_rotational_N6.txt};
\addplot[mesh, line legend, thick, point meta max = 7.5, point meta=explicit]
table[x=t,y=pmodel, meta=pmodel]{data/static_result_strand_rotational_N6.txt};
\addplot[scatter, only marks, scatter/classes={0={bw_6}, 1={bw_5}, 2={bw_4}, 3={bw_3}, 4={bw_2}, 5={bw_1}},
        scatter src=explicit symbolic, mark=*, thick, mark size=0.8pt] 
table[x=t,y=pref, meta=style]{data/static_result_strand_rotational_N6_few.txt};
\addplot[scatter, only marks, scatter/classes={0={bw_6}, 1={bw_5}, 2={bw_4}, 3={bw_3}, 4={bw_2}, 5={bw_1}},
        scatter src=explicit symbolic, mark=x, thick, mark size=2pt] 
table[x=t,y=pmodel, meta=style]{data/static_result_strand_rotational_N6_few.txt};
\legend{Reference, ROHM}
\end{axis}
\end{tikzpicture}
\end{subfigure}
\caption{Comparison of ROHM model results with the reference solution. Rotating transverse magnetic field $\hs^{(3)}(t)$. Chain of \SCs\ with $N=6$ cells and parameters of Fig.~\ref{tab_param_static}. Solutions at selected time instants are indicated by the colored circles (reference solution) and crosses (ROHM model). (Up) Internal magnetic field and magnetization of the filament. The ROHM model curve is colored as a function of the instantaneous power loss. Same legend as in the bottom sub-figure. Values are in tesla (T). (Down) Dissipated power per unit length as a function of time.}
    \label{static_result_rotational}
\end{figure}
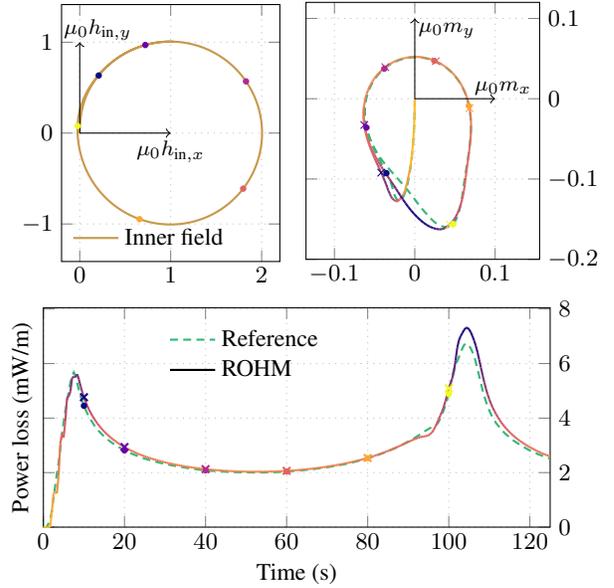

The parameters in Section~\ref{sec_application_static_identification} are defined based on the results of a unidirectional excitation, but all the ROHM model equations are vectorial, and the model is therefore directly applicable to general excitations. We show in Fig.~\ref{static_result_rotational} the results for a bidirectional rotating transverse field excitation defined by:
\begin{align}
\mu_0 \hs^{(3)}(t) =\quad & \mu_0\hmax \sin(\pi ft)\sin(\pi ft)\ex\notag\\
+ &\mu_0\hmax \sin(\pi ft)\cos(\pi ft)\ey,
\end{align}
with $\mu_0\hmax=2$~T, $f=0.01$~Hz, and $\ex$, $\ey$ the unit vectors in $x$-, $y$-directions. The influence of the number of cells $N$ on the relative error is shown in Fig.~\ref{fig_static_N_effect}, a constant error of $\approx 3\%$ remains for large values of $N$.

The ROHM model reproduces well the magnetization and power loss. The angle between the inner field $\h_\text{in}$ and the magnetization vector $\vec m$ is faithfully described, as shown by the solutions at selected time instants in Fig.~\ref{static_result_rotational}. The agreement is not perfect, but very satisfying provided that the model parameters are fully identified using results from a unidirectional situation, with absolutely no information about the strand response under bidirectional excitations. 

\section{Results with a chain of \CSCs}\label{sec_multiStrand}



As a generalization of the previous section, we consider the same 54-filament strand subject to a transverse field, but now with a conducting matrix and higher rates of field changes, so that both magnetization and loss exhibit frequency dependence. The reference solutions are obtained as described in Section~\ref{sec_context} with the CATI method in order to account for coupling current effects due to the twist of the filaments.

We start the analysis by choosing the ROHM model parameters based on reference solutions. We then illustrate the results for wide ranges of field amplitudes and frequencies.

\subsection{Parameter identification}

\begin{figure}[h!]
\begin{center}
\includegraphics[width=\linewidth]{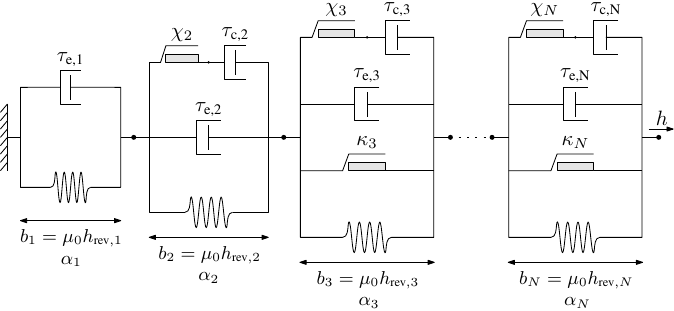}
\caption{Chain of \CSCs\ to model a multifilamentary strand rate-dependent response.}
\label{mechanical_analogy_combined_special}
\end{center}
\end{figure}

\begin{figure}[h!]
\begin{center}
\includegraphics[width=\linewidth]{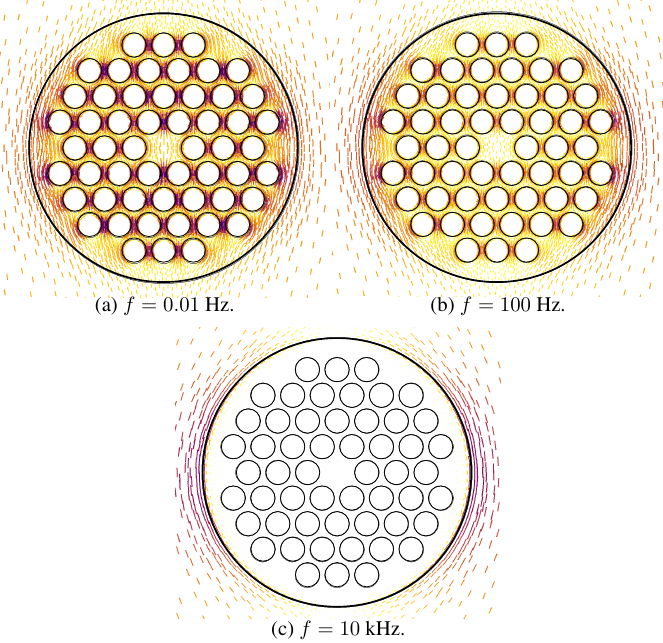}
\caption{Field distribution at low fields ($\mu_0 h_\text{max}= 10$~mT) and different frequencies. The lines are parallel to the local magnetic field, the larger and darker they are, the higher the field amplitude. (a) Uncoupled filaments and limited magnetization. (b) Increased magnetization due to coupled filaments. (c) Almost perfect diamagnetism due to eddy currents.}
\label{low_field_inner_cells}
\end{center}
\end{figure}

\begin{figure}[h!]
    \centering
\centering
\tikzsetnextfilename{lowField_magn}
\begin{tikzpicture}[trim axis left, trim axis right][font=\small]
\pgfplotsset{set layers}
 \begin{axis}[
tick label style={/pgf/number format/fixed},
width=\linewidth,
height=6.5cm,
grid = major,
grid style = dotted,
ymin=-20, 
ymax=1,
xmin=-1, 
xmax=20,
xlabel={$\mu_0\hinS$ (mT)},
ylabel={Magnetization $\mu_0 m$ (mT)},
ylabel style={yshift=-2.8em},
xlabel style={yshift=0.5em},
xticklabel style={yshift=0.1em},
yticklabel style={xshift=0em},
yticklabel pos=right,
legend columns=4,
transpose legend,
legend style={at={(0.01, 0.01)}, cells={anchor=west}, anchor=south west, draw=none,fill opacity=0, text opacity = 1}
]
\addplot[vir_6, thick] 
table[x=mu0h,y=m]{data/lowField_0.01Hz.txt};
    \addplot[vir_5, thick] 
table[x=mu0h,y=m]{data/lowField_0.1Hz.txt};
    \addplot[vir_4, thick] 
table[x=mu0h,y=m]{data/lowField_1.0Hz.txt};
    \addplot[vir_3, thick] 
table[x=mu0h,y=m]{data/lowField_10.0Hz.txt};
\addlegendimage{empty legend}
\addlegendentry{} 
    \addplot[vir_2, thick] 
table[x=mu0h,y=m]{data/lowField_100.0Hz.txt};
    \addplot[vir_1, thick] 
table[x=mu0h,y=m]{data/lowField_1000.0Hz.txt};
    \addplot[vir_0, thick] 
table[x=mu0h,y=m]{data/lowField_10000.0Hz.txt};
\addplot[vir_6, thick, densely dashed] 
table[x=mu0h,y=m]{data/lowField_ref_0.01Hz.txt};
    \addplot[vir_5, thick, densely dashed] 
table[x=mu0h,y=m]{data/lowField_ref_0.1Hz.txt};
    \addplot[vir_4, thick, densely dashed] 
table[x=mu0h,y=m]{data/lowField_ref_1.0Hz.txt};
    \addplot[vir_3, thick, densely dashed] 
table[x=mu0h,y=m]{data/lowField_ref_10.0Hz.txt};
    \addplot[vir_2, thick, densely dashed] 
table[x=mu0h,y=m]{data/lowField_ref_100.0Hz.txt};
    \addplot[vir_1, thick, densely dashed] 
table[x=mu0h,y=m]{data/lowField_ref_1000.0Hz.txt};
    \addplot[vir_0, thick, densely dashed] 
table[x=mu0h,y=m]{data/lowField_ref_10000.0Hz.txt};
\legend{$f=0.01$ Hz, $f=0.1$ Hz, $f=1$ Hz, $f=10$ Hz, \empty, $f=10^{2}$ Hz, $f=10^{3}$ Hz, $f=10^{4}$ Hz}
\end{axis}
\end{tikzpicture}
\vspace{-0.2cm}
\caption{Strand magnetization for $\mu_0 \hmax = 10$~mT at different frequencies, from virgin state at time $t=0$ to time $t = 0.25/f$ (for better clarity). Dashed curves are reference solutions. Solid curves are from the ROHM model.}
    \label{lowField_magn}
\end{figure}
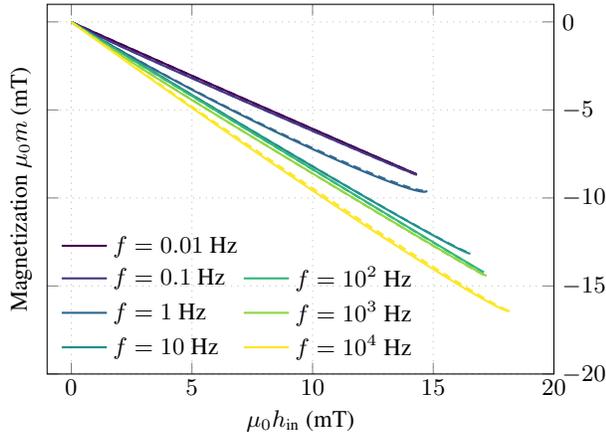

We choose the chain structure represented in Fig.~\ref{mechanical_analogy_combined_special}. Compared to the chain of \SCs\ of the previous section, two cells now have a zero irreversibility parameter, $\kappa_1 = 0$ A/m and $\kappa_2 = 0$ A/m. 
These two cells allow to reproduce the transitions between the three regimes illustrated in Fig.~\ref{low_field_inner_cells} at low fields, associated with different magnetization slopes, as shown in Fig.~\ref{lowField_magn}. The curves are obtained with the parameter values of Table~\ref{tab_param_dynamic}, which are discussed below.

At low frequencies, the dominant contribution to magnetization and loss comes from superconducting filament hysteresis. We can reuse the material parameters found in the previous section for the chain of \SCs\ as is given in Fig.~\ref{tab_param_static}. In order to reproduce losses at fields lower than $150$~mT, we also introduce $8$ new cells with lower irreversibility parameters, logarithmically spaced from $1$~mT to $80$~mT. The associated weights can be identified automatically as proposed in Section~\ref{sec_paramIdentification_static}, or based on analytical solutions at low fields. These additional cells are optional and only needed to reproduce accurate results on wide field amplitude ranges.

We then choose the time constants. A good fit with the reference solution is obtained with the values provided in Table~\ref{tab_param_dynamic}. Larger time constants are chosen for cells associated with higher irreversibility parameters, in order to reproduce the shifts of the peak coupling and eddy loss observed in Fig.~\ref{strand_lossCycle} and discussed in Section~\ref{sec_dynamicResponse}.

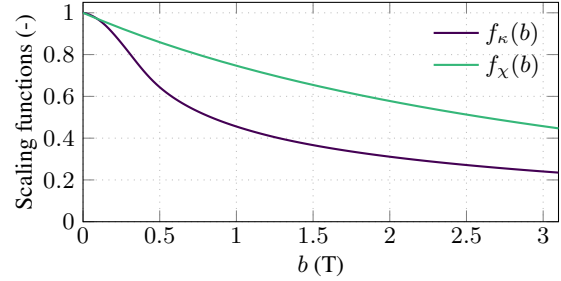
\begin{figure}[!h]
\centering
\tikzsetnextfilename{f_chi}
\begin{tikzpicture}[trim axis left, trim axis right][font=\small]
\pgfplotsset{set layers}
 \begin{axis}[
tick label style={/pgf/number format/fixed},
width=0.95\linewidth,
height=4.5cm,
grid = major,
grid style = dotted,
ymin=0, 
ymax=1.05,
xmin=0, 
xmax=3.1,
domain=0:3.1,
ytick={0, 0.2, 0.4, 0.6, 0.8, 1.0},
xlabel={$b$ (T)},
ylabel={Scaling functions (-)},
ylabel style={yshift=-1.5em},
xlabel style={yshift=0.6em},
xticklabel style={yshift=0.1em},
yticklabel style={xshift=0em},
legend columns=4,
transpose legend,
legend style={at={(0.99, 0.96)}, cells={anchor=west}, anchor=north east, draw=none,fill opacity=0, text opacity = 1}
]
\addplot[vir_6, thick] 
table[x=b,y=fk]{data/f_kappa.txt};
\addplot[vir_2, thick] 
{(1-x/15)/(1+x/4)};
\legend{$f_\kappa(b)$, $f_\chi(b)$}
\end{axis}
\end{tikzpicture}
\caption{Field-dependent scaling functions for the irreversibility parameters $\kappa_k(b)$ and $\chi_k(b)$ for the chain of \CSCs.}
\label{scaling_functions_kappa_chi}
\end{figure}

Finally, we fix the scaling $f_\chi(\b)$ and the parameters $\bar \kc_k$. A decent fit is obtained with the curve in Fig.~\ref{scaling_functions_kappa_chi}. The parameters $\bar \kc_k$ are then tuned manually to obtain the values in Table~\ref{tab_param_dynamic}.


\begin{table}[!h]\small
\centering
\begin{tabular}{r r r l l l}
\hline
\multicolumn{1}{c}{$k$} & \multicolumn{1}{c}{$\mu_0\bar\kappa_k$} & \multicolumn{1}{c}{$\alpha_k$} & \multicolumn{1}{c}{$\tek$} & \multicolumn{1}{c}{$\tck$} & \multicolumn{1}{c}{$\mu_0\bar\chi_k$}\\
\multicolumn{1}{c}{(-)} & \multicolumn{1}{c}{(mT)} & \multicolumn{1}{c}{($\%$)} & \multicolumn{1}{c}{(ms)} & \multicolumn{1}{c}{(s)} & \multicolumn{1}{c}{(T)}\\
\hline
$1$  & $0.0$   & $16.38$ & $0.02$ & $0.00$ & $0.00$\\
$2$  & $0.0$   & $22.62$ & $0.02$ & $0.08$ & $1.13$ \\
\hline
$3$  & $1.0$   & $0.05$  & $0.02$ & $0.08$ & $0.01$\\
$4$  & $1.9$   & $0.10$  & $0.02$ & $0.08$ & $0.02$\\
$5$  & $3.5$   & $0.19$  & $0.02$ & $0.08$ & $0.03$\\
$6$  & $6.5$   & $0.35$  & $0.02$ & $0.08$ & $0.07$\\
$7$  & $12.2$  & $0.66$  & $0.02$ & $0.08$ & $0.09$\\
$8$  & $22.9$  & $1.24$  & $0.02$ & $0.08$ & $0.14$\\
$9$  & $42.9$  & $2.32$  & $0.02$ & $0.08$ & $0.21$\\
$10$ & $80.2$  & $4.33$  & $0.30$ & $0.28$ & $0.40$\\
\hline
$11$ & $150.0$ & $18.07$ & $0.30$ & $0.28$ & $0.75$\\
$12$ & $300.0$ & $16.99$ & $0.30$ & $0.28$ & $2.10$\\
$13$ & $450.0$ & $9.31$  & $0.30$ & $0.28$ & $3.15$\\
$14$ & $600.0$ & $5.16$  & $0.30$ & $0.28$ & $4.20$\\
$15$ & $750.0$ & $2.22$  & $0.30$ & $0.28$ & $5.25$\\
\hline
\end{tabular}
\caption{Parameters of a chain of \CSCs\ with $N = 15$.}
\label{tab_param_dynamic}
\end{table}

\subsection{Results and numerical performance}

\begin{figure}[h!]
    \centering
\centering
\tikzsetnextfilename{magnCurves_1T_comp}
\begin{tikzpicture}[trim axis left, trim axis right][font=\small]
\pgfplotsset{set layers}
 \begin{axis}[
tick label style={/pgf/number format/fixed},
width=\linewidth,
height=7.5cm,
grid = major,
grid style = dotted,
ymin=-2, 
ymax=2,
xmin=-2, 
xmax=2,
xlabel={$\mu_0\hinS$ (T)},
ylabel={Magnetization $\mu_0 m$ (T)},
ylabel style={yshift=-2.8em},
xlabel style={yshift=0.5em},
xticklabel style={yshift=0.1em},
yticklabel style={xshift=0em},
yticklabel pos=right,
legend columns=4,
transpose legend,
legend style={at={(0.01, 0.011)}, cells={anchor=west}, anchor=south west, draw=none,fill opacity=0, text opacity = 1}
]
\addplot[vir_6, thick] 
table[x=mu0h,y=m]{data/hyst_at_1T_0.01Hz.txt};
    \addplot[vir_5, thick] 
table[x=mu0h,y=m]{data/hyst_at_1T_0.1Hz.txt};
    \addplot[vir_4, thick] 
table[x=mu0h,y=m]{data/hyst_at_1T_1.0Hz.txt};
    \addplot[vir_3, thick] 
table[x=mu0h,y=m]{data/hyst_at_1T_10.0Hz.txt};
\addlegendimage{empty legend}
\addlegendentry{} 
    \addplot[vir_2, thick] 
table[x=mu0h,y=m]{data/hyst_at_1T_100.0Hz.txt};
    \addplot[vir_1, thick] 
table[x=mu0h,y=m]{data/hyst_at_1T_1000.0Hz.txt};
    \addplot[vir_0, thick] 
table[x=mu0h,y=m]{data/hyst_at_1T_10000.0Hz.txt};
\addplot[vir_6, thick, densely dashed] 
table[x=mu0h,y=m]{data/ref_at_1T_0.01Hz.txt};
    \addplot[vir_5, thick, densely dashed] 
table[x=mu0h,y=m]{data/ref_at_1T_0.1Hz.txt};
    \addplot[vir_4, thick, densely dashed] 
table[x=mu0h,y=m]{data/ref_at_1T_1.0Hz.txt};
    \addplot[vir_3, thick, densely dashed] 
table[x=mu0h,y=m]{data/ref_at_1T_10.0Hz.txt};
\addlegendimage{empty legend}
\addlegendentry{} 
    \addplot[vir_2, thick, densely dashed] 
table[x=mu0h,y=m]{data/ref_at_1T_100.0Hz.txt};
    \addplot[vir_1, thick, densely dashed] 
table[x=mu0h,y=m]{data/ref_at_1T_1000.0Hz.txt};
    \addplot[vir_0, thick, densely dashed] 
table[x=mu0h,y=m]{data/ref_at_1T_10000.0Hz.txt};
\legend{$f=0.01$ Hz, $f=0.1$ Hz, $f=1$ Hz, $f=10$ Hz, \empty, $f=10^{2}$ Hz, $f=10^{3}$ Hz, $f=10^{4}$ Hz}
\end{axis}
\end{tikzpicture}
\vspace{-0.2cm}
\caption{Strand magnetization for $\mu_0\hmax = 1$~T at different frequencies. Solid curves are results from the ROHM model with parameters of Table~\ref{tab_param_dynamic}. Dashed curves are reference solutions.}
    \label{magnCurves_1T_comp}
\end{figure}
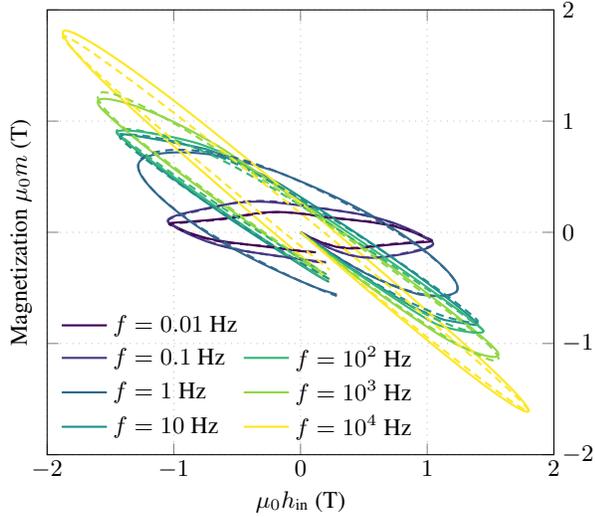

Using the model parameters given in Table~\ref{tab_param_dynamic}, we compare the predictions of the ROHM model with the reference solutions. The loss per cycle as a function of frequency and magnetization loops are given in Figs.~\ref{strand_lossCycle_results} and \ref{magnCurves_1T_comp}.


\begin{figure}[h!]
\centering
 \begin{subfigure}[b]{0.99\linewidth}  
 \centering
\tikzsetnextfilename{strand_lossCycle_results_1T}
\begin{tikzpicture}[trim axis left, trim axis right][font=\small]
\pgfplotsset{set layers}
 	\begin{loglogaxis}[
	tick label style={/pgf/number format/fixed},
    width=\linewidth,
    height=3.8cm,
    grid = major,
    grid style = dotted,
    ymin=2e-3, 
    ymax=2,
    xmin=0.01, 
    xmax=10000,
    xticklabels={},
    ylabel style={yshift=-2.8em},
    xlabel style={yshift=0.5em},
    xticklabel style={yshift=0.1em},
    yticklabel style={xshift=0em},
    yticklabel pos=right,
    legend columns=2,
    legend style={at={(0.28, 0.06)}, cells={anchor=west}, anchor=south, draw=none,fill opacity=0, text opacity = 1, legend image code/.code={\draw[##1,line width=1pt] plot coordinates {(0cm,0cm) (0.3cm,0cm)};}}
    ]
    \addplot[black, thick] 
    table[x=f,y=tot_model]{data/strand_lossCycle_1T.txt};
        \addplot[vir_3, thick] 
    table[x=f,y=fil_model]{data/strand_lossCycle_1T.txt};
        \addplot[vir_0, thick] 
    table[x=f,y=coupling_model]{data/strand_lossCycle_1T.txt};
        \addplot[myorange, thick] 
    table[x=f,y=eddy_model]{data/strand_lossCycle_1T.txt};
        \addplot[black, thick, densely dashed] 
    table[x=f,y=tot_ref]{data/strand_lossCycle_ref_1T.txt};
        \addplot[vir_3, thick, densely dashed] 
    table[x=f,y=fil_ref]{data/strand_lossCycle_ref_1T.txt};
        \addplot[vir_0, thick, densely dashed] 
    table[x=f,y=coupling_ref]{data/strand_lossCycle_ref_1T.txt};
        \addplot[myorange, thick, densely dashed] 
    table[x=f,y=eddy_ref]{data/strand_lossCycle_ref_1T.txt};
    \legend{Total, Hysteresis, Coupling, Eddy}
\node[anchor=south] at (axis cs: 1000, 0.015) {$\mu_0\hmax = 1$ T};
    \end{loglogaxis}
\end{tikzpicture}%
\end{subfigure}
        \hfill\vspace{-0.4cm}
 \begin{subfigure}[b]{0.99\linewidth}  
 \centering
\tikzsetnextfilename{strand_lossCycle_results_0p1T}
\begin{tikzpicture}[trim axis left, trim axis right][font=\small]
\pgfplotsset{set layers}
 	\begin{loglogaxis}[
	tick label style={/pgf/number format/fixed},
    width=\linewidth,
    height=3.8cm,
    grid = major,
    grid style = dotted,
    ymin=2e-5, 
    ymax=2e-2,
    xmin=0.01, 
    xmax=10000,
    xticklabels={},
    ylabel={Loss per cycle (J/m)},
    ylabel style={yshift=-2.2em},
    xlabel style={yshift=0.5em},
    xticklabel style={yshift=0.1em},
    yticklabel style={xshift=0em},
    yticklabel pos=right,
    legend columns=2,
    legend style={at={(0.35, 0.0)}, cells={anchor=west}, anchor=south, draw=none,fill opacity=0, text opacity = 1, legend image code/.code={\draw[##1,line width=1pt] plot coordinates {(0cm,0cm) (0.3cm,0cm)};}}
    ]
    \addplot[black, thick] 
    table[x=f,y=tot_model]{data/strand_lossCycle_0p1T.txt};
        \addplot[vir_3, thick] 
    table[x=f,y=fil_model]{data/strand_lossCycle_0p1T.txt};
        \addplot[vir_0, thick] 
    table[x=f,y=coupling_model]{data/strand_lossCycle_0p1T.txt};
        \addplot[myorange, thick] 
    table[x=f,y=eddy_model]{data/strand_lossCycle_0p1T.txt};
        \addplot[black, thick, densely dashed] 
    table[x=f,y=tot_ref]{data/strand_lossCycle_ref_0p1T.txt};
        \addplot[vir_3, thick, densely dashed] 
    table[x=f,y=fil_ref]{data/strand_lossCycle_ref_0p1T.txt};
        \addplot[vir_0, thick, densely dashed] 
    table[x=f,y=coupling_ref]{data/strand_lossCycle_ref_0p1T.txt};
        \addplot[myorange, thick, densely dashed] 
    table[x=f,y=eddy_ref]{data/strand_lossCycle_ref_0p1T.txt};
    \node[anchor=south] at (axis cs: 200, 0.002) {$\mu_0\hmax = 0.1$ T};
    \end{loglogaxis}
\end{tikzpicture}%
\end{subfigure}
        \hfill\vspace{-0.2cm}
 \begin{subfigure}[b]{0.99\linewidth}  
        \centering
\tikzsetnextfilename{strand_lossCycle_results_0p01T}
\begin{tikzpicture}[trim axis left, trim axis right][font=\small]
\pgfplotsset{set layers}
 	\begin{loglogaxis}[
	tick label style={/pgf/number format/fixed},
    width=\linewidth,
    height=3.8cm,
    grid = major,
    grid style = dotted,
    ymin=2e-7, 
    ymax=2e-4,
    xmin=0.01, 
    xmax=10000,
	xlabel={Frequency $f$ (Hz)},
    ylabel style={yshift=-2.8em},
    xlabel style={yshift=0.5em},
    xticklabel style={yshift=0.1em},
    yticklabel style={xshift=0em},
    yticklabel pos=right,
    legend columns=2,
    legend style={at={(0.38, 0.03)}, cells={anchor=west}, anchor=south, draw=none,fill opacity=0, text opacity = 1, legend image code/.code={\draw[##1,line width=1pt] plot coordinates {(0cm,0cm) (0.3cm,0cm)};}}
    ]
    \addplot[black, thick] 
    table[x=f,y=tot_model]{data/strand_lossCycle_0p01T.txt};
        \addplot[vir_3, thick] 
    table[x=f,y=fil_model]{data/strand_lossCycle_0p01T.txt};
        \addplot[vir_0, thick] 
    table[x=f,y=coupling_model]{data/strand_lossCycle_0p01T.txt};
        \addplot[myorange, thick] 
    table[x=f,y=eddy_model]{data/strand_lossCycle_0p01T.txt};
        \addplot[black, thick, densely dashed] 
    table[x=f,y=tot_ref]{data/strand_lossCycle_ref_0p01T.txt};
        \addplot[vir_3, thick, densely dashed] 
    table[x=f,y=fil_ref]{data/strand_lossCycle_ref_0p01T.txt};
        \addplot[vir_0, thick, densely dashed] 
    table[x=f,y=coupling_ref]{data/strand_lossCycle_ref_0p01T.txt};
        \addplot[myorange, thick, densely dashed] 
    table[x=f,y=eddy_ref]{data/strand_lossCycle_ref_0p01T.txt};
\node[anchor=south] at (axis cs: 200, 0.00002) {$\mu_0\hmax = 0.01$ T};
    \end{loglogaxis}
\end{tikzpicture}%
\end{subfigure}
 \hfill
\vspace{-0.2cm}
\caption{Energy loss per cycle for three different field amplitudes as a function of frequency.  Solid curves are results from the ROHM model with parameters of Table~\ref{tab_param_dynamic}. Dashed curves are reference solutions. The legend is the same for the three subfigures.}
        \label{strand_lossCycle_results}
\end{figure}
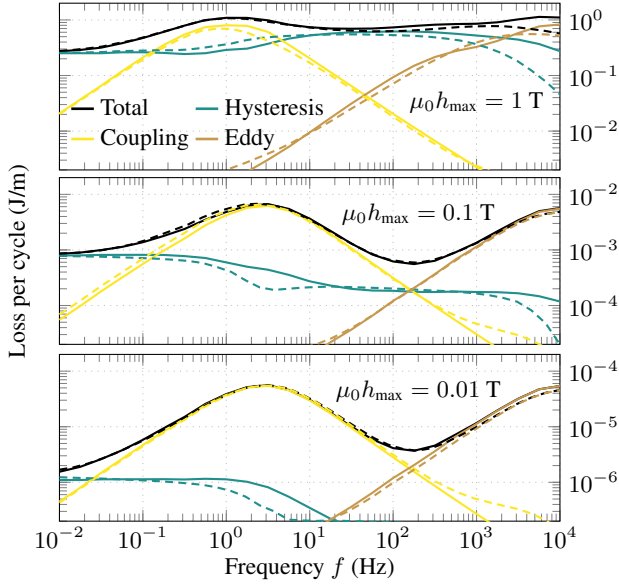

The overall agreement on the total loss and magnetization is very good. The ROHM model correctly reproduces the different regimes of distinct dominant loss contributions and describes the magnetization faithfully. The difference with the reference solution is larger at high frequency and high fields, as can be seen in Fig.~\ref{strand_lossCycle_results}. This can be improved with an adapted choice of model parameters, but as discussed in Section~\ref{sec_cell_coupling}, the matching will never be perfect in the limit $f\to \infty$ as the \CSCs\ do not reproduce a $1/\sqrt{f}$ decrease of the loss per cycle, but rather a $1/f$ decrease. For not too high frequencies however (skin depth not too small compared to strand diameter), the importance of this effect is limited.

As in the rate-independent case, the major advantage of the ROHM model compared to the detailed simulations is the very small associated computational work. Even with $N=15$ cells, each ROHM model simulation is conducted in a few milliseconds, whereas the computational time for a reference solution with the CATI method is of the order of one hour (with $28\,000$ degrees of freedom). The ROHM model enables the simulation of large-scale superconducting systems involving hundreds or thousands of strands, which would be completely unattainable with detailed strand models.



\begin{figure}[h!]
\begin{center}
\includegraphics[width=\linewidth]{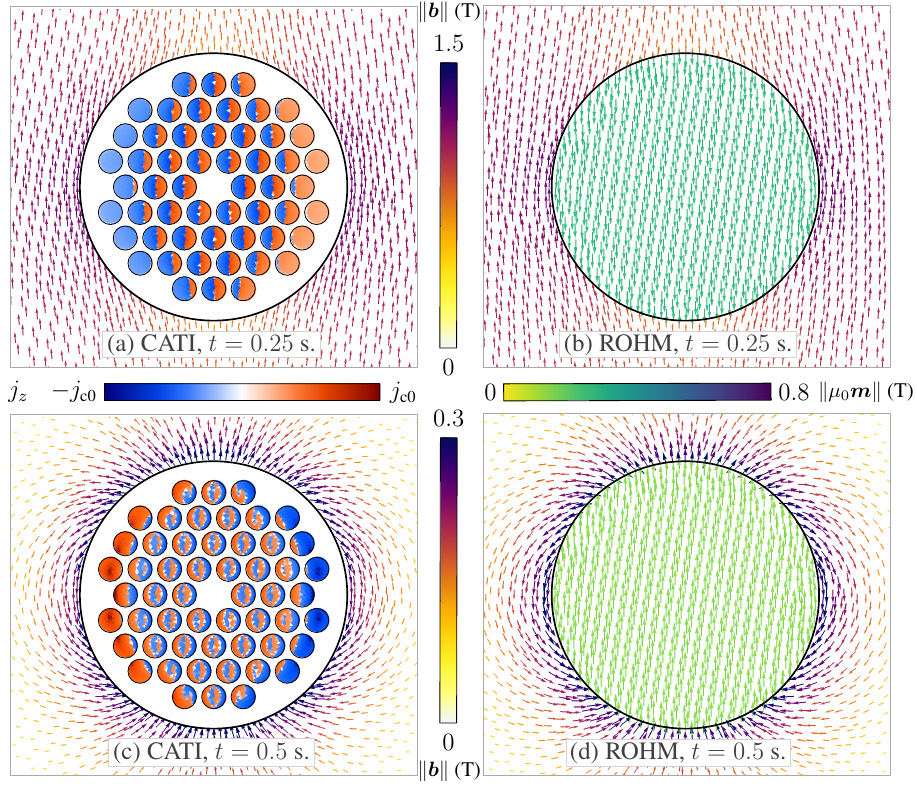}
\caption{Solutions of the reference CATI model and the ROHM model for $\mu_0\hmax = 1$~T and $f=1$~Hz. (a-b) Solutions at $t=0.25$~s. (c-d) Solutions at $t=0.5$~s. Arrows outside of the strand represent the magnetic flux density $\b$ (shared color scale in the middle). Colored elements in (a) and (c) represent the $z$-component of the current density $\j$ (color scale on the left, with $j_\text{c0} = 3\times 10^{10}$ A/m$^2$). Arrows inside the strand in (b) and (d) represent the magnetization $\mu_0 \m = \b - \mu_0 \h$ (color scale on the right).}
\label{homogenization_in_FE}
\end{center}
\end{figure}

As described in Section~\ref{sec_fem}, the ROHM model can also easily be included in a FE model as a homogenized magnetic property for the local fields $\h$ and $\b = \bhyst(\h)$. An illustration of the obtained field is given in Fig.~\ref{homogenization_in_FE} in the simple case of a single strand surrounded by air, obtained with the $\phi$-formulation. The ROHM model reproduces the magnetization of the detailed strand model, such that the field seen outside of the homogenized strand is identical to that seen outside of the detailed strand.

In superconducting magnets, many strands are combined together into cables, which themselves are wound around the magnet aperture. The ROHM model can be used to homogenize the large number of strands into an equivalent homogeneous material. In such a case, the reference detailed solution should be that of a close-packed set of strands, which can typically be modelled via appropriate periodic boundary conditions~\cite{meunier2010homogenization}, rather than a single strand in air, as is done here for the sake of illustration. This may result in different model parameters, but the general approach remains the same.

\section{Conclusion and outlook}

In this work, we introduced the Reduced Order Hysteretic Magnetization (ROHM) model to represent the magnetization and instantaneous power loss induced in composite superconductors by transient transverse magnetic fields. This model is designed by adapting the state-of-the-art energy-based model for ferromagnetic hysteresis~\cite{henrotte2006energy, henrotte2006dynamical}.

We focused specifically on the case of a multifilamentary superconducting strand and proposed two variations of the ROHM model: a rate-independent model, describing hysteresis in superconducting filaments, and a rate-dependent model, accounting for filament coupling, coupling current, and eddy current effects. We proposed a parameter identification approach for both variations of the model, based on reference solutions obtained with detailed strand models, and we described their inclusion in a finite element framework. Finally, we demonstrated that the ROHM model has the potential to reduce the computational cost compared to conventional simulations by several orders of magnitude. The model accuracy can be tailored for specific applications: a very good accuracy (typically with a few percent error on the power loss) in a wide range of field amplitudes and rates of change can be obtained if required. It also allows to represent deformed strands or strands with defects, provided that reference solutions are used in order to fix adapted ROHM model parameters.

This work constitutes a first step towards the homogenization of superconducting magnet windings, in view of enabling the simulation of magneto-thermal transients in superconducting magnet cross sections~\cite{schnaubelt2025transient}, accounting for all loss contributions. Replacing the cables with homogenized materials allows to describe the magnet response accurately and efficiently without computing the detailed current density distribution.

The ROHM model accounts for magnetization and loss induced by an external transverse field. One next step consists in defining a second reduced order model to describe the inductance and power loss induced by transport current. The response of a composite strand to transport current is also hysteretic and rate-dependent, therefore, this second reduced order model can be built using similar concepts as those of the ROHM model. Additionally, the ROHM model and the reduced order model for transport current must be coupled. Indeed, the presence of a transport current affects the magnetization, and vice-versa, the presence of a background magnetic field affects the transport current response. Temperature dependence must also be included in the model parameters for magneto-thermal simulations.

For low-temperature superconducting magnets, interstrand coupling current effects in Rutherford cables must also be accounted for. They can be studied using detailed cable models such as in~\cite{dular2024simulation} combined with homogenized strand models. The homogenization of these effects into reduced order models will be considered as another next step.

The extension of the approach to high-temperature superconducting (HTS) tapes, stacks of HTS tapes, or HTS bulks in view of their homogenization requires to account for the strong anisotropy of these systems. In principle, as the ROHM model equations are all vectorial, this can be done via the definition of direction-dependent irreversibility parameters and time constants. Further study is required in that direction.


\appendix
\section{Magnetization in 2D Problems}\label{app_magnetization}

In this section, we justify the introduction of a factor $2$ in the magnetization Eq.~\eqref{eq_magnetization}, in the case of an infinitely long problem solved in 2D with axial currents.

\begin{figure}[h!]
\begin{center}
\includegraphics[width=\linewidth]{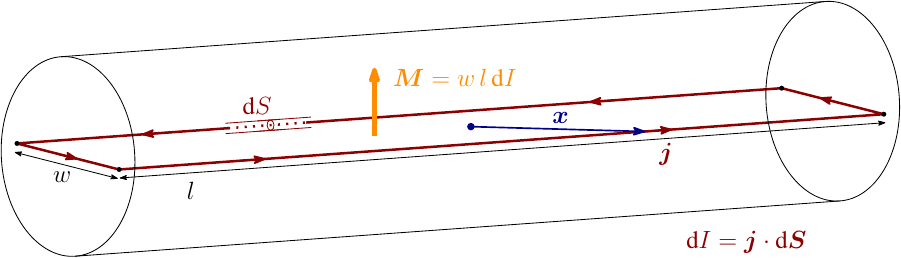}
\caption{Rectangular current loop in a cylinder.}
\label{cylinder_magnetization}
\end{center}
\end{figure}

The magnetic dipole moment $\vec M$ (A\,m$^2$) of a current density distribution $\j$ in a given volume $V$ is evaluated as
\begin{align}
\vec M = \frac{1}{2}\int_{V} \vec x \times \j\, \text{d}V,
\end{align}
with $\vec x$ the position vector. Consider a rectangular current loop of cross section $\text{d}S$, width $w$, and length $l$, carrying a current $\text{d}I = \j \cdot \text{d}\vec S$, as illustrated in Fig.~\ref{cylinder_magnetization}. Accounting for the four sides of the rectangle, the contribution of this current loop to the total magnetic dipole moment is $wl\,\text{d}I$. The two long sides of length $l$ contribute to half of it: $wl\,\text{d}I/2$. The two short sides of length $w$, closing the current loop, contribute to exactly the other half of it: $wl\,\text{d}I/2$.

When the geometry is sufficiently long for end effects to be neglected, the problem is usually solved in 2D, per unit length, on a cross section of the conducting cylinder. To compute the magnetic dipole moment in such a case, however, one must still account for the fact that all current loops must close at the end of the cylinder. These closing currents are not part of the 2D solution, and are therefore missing from any calculation performed solely on the 2D cross section. Because the closing currents conveniently contribute equally to the magnetic dipole moment compared to the other currents, one can account for them by simply introducing a factor $2$, as in Eq.~\eqref{eq_magnetization}.

\section{Applied Field and Internal Field}\label{app_app_and_in_fields}

The energy loss per cycle $Q_\text{tot}$ given by Eq.~\eqref{eq_lossPerCycle_detailed} can be equivalently computed in terms of $\hin = \hs - \m/2$. Indeed, as also shown in~\cite{goldfarb1986internal},
\begin{align}
Q_\text{tot} &= a \oint \mu_0 \hs \cdot \text{d}\vec m\notag\\
&= a \oint \mu_0 (\hin + \m/2) \cdot \text{d}\vec m\notag\\
&= a \oint \mu_0 \hin \cdot \text{d}\vec m + \frac{a\mu_0}{2} \underbrace{\oint  \m \cdot \text{d}\vec m}_{=0\text{ if closed}}\notag\\
&= a \oint \mu_0 \hin \cdot \text{d}\vec m.
\end{align}
In particular, this justifies that the areas enclosed by the loops represented in Figs.~\ref{magnCurves_2T} and \ref{magnCurves_2T_hin} are identical.

\section{ROHM model Jacobian}\label{app_jacobian}

The inclusion of a ROHM model into a FE model makes the problem nonlinear. Equations must then be solved iteratively. In order to implement the Newton-Raphson technique, the Jacobian of the hysteresis law $\b = \boldsymbol{\mathcal{B}}(\h)$ must be evaluated. It is a tensor of order two which reads as follows, using Eq.~\eqref{eq_chainEqn}:
\begin{align}\label{eq_jacobian_firstStep}
\der{\b}{\h} = \sum_{k=1}^N \alpha_k \mu_0 \der{\hrevk}{\h}.
\end{align}
Its expression can be computed independently for each cell. Below, we give its expression for the simple case of a \SC\ and we then generalize for a \CSC. We drop the index $k$ for conciseness.

\subsection{\SC}

For a \SC, the reversible field $\hrev$ is updated with Eq.~\eqref{eq_vpm}. The derivative of this update rule is given by
\begin{align}\label{eq_deriv_vpm}
\der{\hrev}{\h} &= \left\{\begin{aligned}
&\mat 0,\quad \text{if } \|\h - \hrevp\|\le \ku,\\
&\paren{1 - \frac{\ku}{\|\h - \hrevp\|}} \mat I  + \frac{\ku}{\|\h - \hrevp\|^3}\mat K,\\
& \qquad  \text{otherwise,}\notag
\end{aligned}\right.\\
&= \boldsymbol{\mathcal W}_{\ku}(\h, \hrevp),
\end{align}
with $\mat I$ the second-order identity tensor and $\mat K$ the dyadic product $\mat K = (\h - \hrevp)\otimes (\h - \hrevp)$.

\subsection{\CSC}

For a \CSC, we first compute $\hcoupling^\text{trial}$ using Eq.~\eqref{eq_hcoupling_trial}. In the case $\|\hcoupling^\text{trial}\| \le \kc$, using Eqs.~\eqref{eq_hrev_for_dyn}, \eqref{eq_coupling_if_smaller}, and~\eqref{eq_eddy_if_smaller}, we have the reversible field given by
\begin{align}
\hrev = \frac{\Delta t}{\Delta t + \teddy + \tcoupling} \paren{\vec g + \frac{\teddy + \tcoupling}{\Delta t} \hrevp}
\end{align}
whose derivative reads
\begin{align}\label{eq_deriv_vpm_dyn_smaller}
\der{\hrev}{\h} &= \frac{\Delta t}{\Delta t + \teddy + \tcoupling} \der{\vec g}{\h}\notag\\
&=\frac{\Delta t}{\Delta t + \teddy + \tcoupling} \tilde{\boldsymbol{\mathcal W}}_{\ku}(\h, \hrevp, \vec g_\text{(p)}),
\end{align}
where $\tilde{\boldsymbol{\mathcal W}}_{\ku}$ is a second-order tensor similar to that defined in Eq.~\eqref{eq_deriv_vpm}, but with $\|\h - \vec g_\text{(p)}\|\le \ku$ as an additional necessary condition for the first case, as in Eq.~\eqref{eq_update_g}.

By contrast, if $\|\hcoupling^\text{trial}\| > \kc$, one must use Eqs.~\eqref{eq_hrev_for_dyn}, \eqref{eq_coupling_if_larger}, and~\eqref{eq_eddy_if_larger}, which yield the following reversible field
\begin{align}
\hrev = \frac{\Delta t}{\Delta t + \teddy}\paren{\vec g - \kc\frac{\vec g - \hrevp}{\|\vec g - \hrevp\|} + \frac{\teddy}{\Delta t} \hrevp}
\end{align}
whose derivative reads
\begin{align}\label{eq_deriv_vpm_dyn_larger}
\der{\hrev}{\h} &= \frac{\Delta t}{\Delta t + \teddy} \paren{\der{\vec g}{\h} - \der{}{\vec g}\paren{\kc\frac{\vec g - \hrevp}{\|\vec g - \hrevp\|}} \cdot\der{\vec g}{\h}}\notag\\
&= \frac{\Delta t}{\Delta t + \teddy} \boldsymbol{\mathcal W}_{\kc}(\vec g, \hrevp)\cdot \tilde{\boldsymbol{\mathcal W}}_{\ku}(\h, \hrevp, \vec g_\text{(p)}),
\end{align}
where $\boldsymbol{\mathcal W}_{\kc}$ is the second-order tensor defined by
\begin{align}
\boldsymbol{\mathcal W}_{\kc}(\vec g, \hrevp) = \paren{1 - \frac{\kc}{\|\vec g - \hrevp\|}} \mat I&\notag \\
+ \frac{\kc}{\|\vec g - \hrevp\|^3} \mat L,&
\end{align}
with $\mat L = (\vec g - \hrevp)\otimes (\vec g - \hrevp)$.

\section*{References}
\bibliographystyle{ieeetr}
\bibliography{../paperReferences}

\end{document}